\documentclass[12pt]{article}
\usepackage[numbers]{natbib}
\usepackage{a4}
\usepackage{amsmath}
\usepackage{amssymb}
\usepackage{latexsym}
\usepackage{epsfig}
\renewcommand{\theequation}{\arabic{section}.\arabic{equation}}
\def \ii{{\mathrm{i}}}
\def \d{{\mathrm{d}}}

\def \pd{\partial}

\def \e{{\mathrm{e}}}

\def \tl#1{\overset{\kern 1pt\circ}{#1}}
\def \TL#1{\overset{\kern -3pt \circ}{#1}}
\def \TLL#1{\overset{\kern -7pt \circ}{#1}}

\def \Bbeta{\boldsymbol{\beta}}
\def \Be{\boldsymbol{e}}

\def \rr{{\boldsymbol{r}}}
\def \qq{{\boldsymbol{q}}}

\begin{document}
\title{{\bf The gauge theory of dislocations: static solutions of screw and edge dislocations}}
\author{
Markus Lazar$^\text{}$\footnote{E-mail: lazar@fkp.tu-darmstadt.de (M.~Lazar).} ,
Charalampos Anastassiadis,
\\ \\
${}^\text{}$
        Emmy Noether Research Group,\\
        Department of Physics,\\
        Darmstadt University of Technology,\\
        Hochschulstr. 6,\\
        D-64289 Darmstadt, Germany
}

\date{\today}
\maketitle


\begin{abstract}
We investigate the $T(3)$-gauge theory of static dislocations in continuous solids.
We use the most general linear constitutive relations 
bilinear in the elastic distortion tensor and dislocation density tensor
for the force and pseudomoment stresses of an isotropic solid. 
The constitutive relations contain six material parameters. 
In this theory both the force and pseudomoment stresses are asymmetric. 
The theory possesses four characteristic lengths 
$\ell_1$, $\ell_2$, $\ell_3$ and $\ell_4$
which are given explicitely.
We first derive the three-dimensional Green tensor of the master equation 
for the force stresses in the translational gauge theory of dislocations.
We then investigate the situation of generalized plane strain (anti-plane strain 
and plane strain).
Using the stress function method, we find modified stress functions
for screw and edge dislocations. 
The solution of the screw dislocation is given in terms of one independent
length $\ell_1=\ell_4$. 
For the problem of an edge dislocation,
only two characteristic lengths $\ell_2$ and $\ell_3$ arise
with one of them being the same $\ell_2=\ell_1$ as for the screw dislocation. 
Thus, this theory possesses only two independent lengths for generalized plane strain.
If the two lengths $\ell_2$ and $\ell_3$ of an edge dislocation are equal, we obtain
an edge dislocation which is the gauge theoretical version 
of a modified Volterra edge dislocation. 
In the case of symmetric stresses we recover well known results obtained earlier.
\\

\end{abstract}

\section{Introduction}
\setcounter{equation}{0}
\label{intro}
\subsection{Reviewing remarks}
Gauge theories are very successful in high energy physics to understand and
to explain fundamental interactions (see, e.g., ~\citep{Rai,Rubakov}).
Important solutions in gauge theories are topological strings
like the so-called Abrikosov-Nielson-Olesen string~\citep{Abrikosov,NO}.
It is also known that the theories of gravity and generalized gravity can be understood
as gauge theories of the Poincar{\'e} group and the affine group, respectively~\citep{Hehl76,Hehl95}. 
For example, the teleparallel formulation of Einstein's general relativity 
can be understood as a gauge theory of the four-dimensional translation group
(see, e.g., \citet{Hehl07}). Thus, the torsion tensor is not just a tensor, but rather a very 
special tensor that is related to the translation group~\citep{Hehl95}.
In such theories mathematical solutions for cosmic strings and other topological defects
in space-time are found~\citep{Vilenkin}.
But it is hard to observe them by experiment.

An application of the three-dimensional translational gauge theory is the 
gauge theoretical formulation of dislocations.
The torsion tensor plays the role as the translational field strength.
In such a theory dislocations appear quite naturally as topological defects 
(strings).
Dislocations play an important role in solid state physics~\citep{HL}.
It is easy to investigate them experimentally.
Therefore, the gauge theory of dislocations is the best example to 
investigate the translational gauge theory as a physical field theory.
Pioneering work in the field of the translational gauge theory of defects in 
solids was made by~\citet{Edelen81,Edelen83,Edelen88}.
It is well known that dislocations are the fundamental carrier of plasticity.
The hope of the gauge theory of dislocations is to understand plasticity as 
a `fundamental' interaction in solids.
Materials display strong size effects when the characteristic length scales 
are of the order of micrometers down to a few nanometers.
With the interest in miniaturization of devices, the length scales associated with  
nanodevices are very small and classical theories of elasticity and plasticity 
are not applicable. 
Classical elasticity and plasticity theories cannot explain the size dependence 
because their constitutive laws possess no internal material lengths.
In metals, the physical origin of size effects can be related to dislocations.
Gauge theory of dislocations may be used to explain size effects and to 
model mechanical properties of miniaturized devices.
In the gauge theory of dislocations the dislocation density tensor which is the curl of the elastic distortion or of the negative plastic distortion 
multiplied by material parameters which define internal material lengths
enter the constitutive relations.
The gauge theory of dislocations shall be a step in the direction of a microscopic
theory of plasticity that incorporates interaction of structural defects.
The dislocation should play a similar role as a gauge boson 
like the photon as a gauge boson in Maxwell's theory.
Nevertheless, the theory of~\citet{Edelen83,Edelen88} has some lacks.
They used very special constitutive relations. 
For example, the force stress tensor is symmetric.
But in the continuum theory of dislocations it is known that the force
stress tensor has to be asymmetric at least in the dislocation core region~\citep{HK65,Kroener69} and that moment stress is the specific response to dislocations~\citep{Kroener92}. Such a moment stress cannot be calculated from elastic moduli only.
A realistic constitutive law between the moment stress and the dislocation density tensor has to be used to built a physical dislocation theory.

Therefore, such a theory is not the 
most general isotropic dislocation gauge theory. 
Their theory is just able to give an acceptable solution of a screw dislocation after
some ad hoc assumptions~\citep{VS88,Edelen96}. 
Moreover, it is not possible to give a correct solution of
an edge dislocation. Especially, the stress component $\sigma_{zz}$ is incorrect
and the condition of plane strain is not satisfied~\citep{Malyshev00}.
Such a dislocation gauge theory with symmetric force stresses
possesses only one additional material parameter
and two length scales can be defined to describe size-effects.
In order to explain bending and torsion experiments one needs
at least two length scales.
One question that arises is how many characteristic length scales 
should appear in an improved gauge theory of dislocations?
Recently, \citet{Sharma05} have combined Edelen's gauge model with a simple strain
gradient elasticity. These authors applied it to a screw dislocation.

These drawbacks of the Edelen model have been a motivation for further investigations~\citep{Malyshev00,Lazar00}. 
The so-called Einstein choice has been investigated by~\citet{Malyshev00} and 
\citet{Lazar02a,Lazar02b}.
Their solution of a screw dislocation looks to be acceptable for theories with
symmetric force stresses.
In the solution of an edge dislocation found by~\citet{Malyshev00}, a strange
far field emerges, which shows an oscillating behavior.
In addition, the condition of plane strain is not fulfilled.
Later, \citet{Lazar03} tried to find a more realistic gauge theoretical solution
for an edge dislocation. 
He used another constitutive relation and it was necessary to
introduce additional gradients of the elastic rotation which is a little sophisticated. 
Thus, up to now, the case of an edge dislocation 
has not been treated satisfactory in the gauge theoretical framework.
Quite recently,~\citet{Malyshev07} has used the Einsteinian $T(3)$-gauge approach to calculate
the stress tensor of a screw dislocation in the second order.
Some other proposals are given by~\citet{KV92,PK01,Kroener01}.
We want to note that all these kinds of gauge theories of dislocations 
use symmetric force stresses and some special choices for the constitutive law
of the moment (or hyper) stresses. 

\subsection{The subject of the paper}

In this paper we want to investigate the translational gauge theory of dislocations 
with the most general linear isotropic constitutive relations 
quadratic in the physical state quantities 
for the force and 
pseudomoment stress tensors with six material coefficients.
Such a theory was originally proposed by~\citet{Lazar02a}. However,
\citet{Lazar02a} used a simplified version for dislocations.
Our purpose is to derive the general framework of such a theory
and to apply it to find new analytical solutions of straight screw and edge
dislocations
in an infinite medium.
The paper consists of five sections.
Section~\ref{intro} is the introduction.
In section~\ref{Basics} we give the basics of the translational
gauge theory of dislocations.
We give the relation for positive definiteness of the elastic energy and the 
dislocation core energy.
We give the relation between the pseudomoment stress
tensor and the moment stress tensor.
We derive the Euler-Lagrange equation in terms of the asymmetric stress
tensor.
This is the central equation for an isotropic material with dislocations.
We set up the relation to other gauge theories of dislocations proposed earlier. 
We calculate the three-dimensional Green tensor for the central equation of 
the asymmetric force stresses.
In this manner we are able to introduce four characteristic length scales
of the gauge theory with asymmetric force stresses.
Later we investigate dislocations in the framework of generalized plane strain.
Sections~\ref{screw} and \ref{edge} 
present the gauge theoretical solutions of straight screw and edge dislocations
in an infinite medium with asymmetric force stresses. 
We will calculate the elastic distortion and dislocation density (torsion)
tensors and the force and pseudomoment stresses of screw and edge dislocations.
We prove that the force stresses of a screw dislocation and an edge
dislocation have to be asymmetric instead of symmetric
in order to guarantee nonnegative energy.
Section~\ref{concl} concludes the paper. 
Details of dislocations in asymmetric elasticity are given in
the appendices~\ref{appendixA} and \ref{appendixB}.

\section{Gauge theory of dislocations}
\setcounter{equation}{0}
\label{Basics}
\subsection{Foundations}

In the translational gauge theory of dislocations the displacement vector
$u_i$ is defined up to a gauge transformation -- a local translation $\tau_i(x)$
\begin{align}
\label{}
u_i\rightarrow u_i+\tau_i(x)
\end{align}
and the translational gauge field $\phi_{ij}$
transforming under the local translation in a suitable form
\begin{align}
\phi_{ij}\rightarrow \phi_{ij}-\tau_{i,j}(x).
\end{align}
From the mathematical point of view, the gauge field $\phi_{ij}$
is the translational part of the generalized affine connection~\citep{Lazar00}.
In the translational gauge theory, the elastic distortion is
defined by the translational gauge-covariant derivative according to
\begin{align}
\beta_{ij}=\nabla_j u_i:=u_{i,j}+\phi_{ij}.
\end{align}
Thus, the elastic distortion is gauge independent and, therefore, a physical
state quantity.
The asymmetric tensor $\phi_{ij}$ may be identified as the negative plastic distortion
tensor~\citep{Lazar00,Lazar02a}.
The elastic strain energy is a function of the elastic distortion.
It reads
\begin{align}
W_{\text{el}}=\frac{1}{2}\,\beta_{ij}\sigma_{ij}
\end{align}
with the specific response
\begin{align}
\label{ST}
\sigma_{ij}=\frac{\pd W_{\text{el}}}{\pd \beta_{ij}}.
\end{align}
It is the analog of the Piola-Kirchhoff stress tensor. It is an
asymmetric stress tensor. The most general isotropic constitutive
relation has three material coefficients. Thus, the asymmetric
force stress tensor has the form\footnote{We are using
the notations $A_{(ij)}\equiv\frac{1}{2}(A_{ij}+A_{ji})$ and
$A_{[ij]}\equiv\frac{1}{2}(A_{ij}-A_{ji})$.}
\begin{align}
\label{stress}
\sigma_{ij}=\lambda\, \delta_{ij}\beta_{kk}+2\mu\, \beta_{(ij)}+2\gamma\, \beta_{[ij]}.
\end{align}
Here $\mu$ and $\lambda$ are the Lam{\'e} coefficients.
The coefficient $\gamma$ is an additional material parameter due to the skew-symmetric
part of the elastic distortion (the elastic rotation). 
Thus, $\gamma$ is the module of rotation (see also~\citet{Kroener76}).
The skew-symmetric stress $\sigma_{[ij]}$ is caused by the (local) elastic
distortion $\beta_{[ij]}$.
If $\gamma=0$ in Eq.~(\ref{stress}), we obtain the Hooke law
with $\sigma_{[ij]}=0$.
Thus, in this case the elastic rotation $\beta_{[ij]}$ is undetermined.  
The positive semi-definiteness of the elastic distortion energy 
$W_{\text{el}}\ge 0$ requires the restriction
\begin{align}
\label{PSD1}
\mu\ge 0,\qquad\gamma \ge 0,\qquad 2\mu+3\lambda\ge 0 .
\end{align}

Another physical state quantity is the translational gauge field strength -- the torsion tensor
\begin{align}
\label{Tor0}
T_{ijk}=\phi_{ik,j}-\phi_{ij,k},
\qquad T_{ijk}=-T_{ikj}
\end{align}
and
\begin{align}
\label{Tor}
T_{ijk}=\beta_{ik,j}-\beta_{ij,k}.
\end{align}
It has the physical meaning of the dislocation density tensor
\begin{align}
\alpha_{ij}=\frac{1}{2}\, \epsilon_{jkl}T_{ikl}=\epsilon_{jkl}\beta_{il,k}.
\end{align}
They fulfill the translational Bianchi identity
\begin{align}
\label{BI} \epsilon_{jkl} T_{ijk,l}=0,\qquad \alpha_{ij,j}=0.
\end{align}
Thus, the physical state quantities are the elastic distortion, $\beta_{ij}$,
and the dislocation density tensor $T_{ijk}$.

The dislocation core energy contains only the translational gauge field strength
and is given by
\begin{align}
\label{W-core}
 W_{\rm disl}=
\frac{1}{4}\, T_{ijk} H_{ijk}
\end{align}
with the translational gauge field momentum tensor
\begin{align}
\label{moment}
H_{ijk}=2\frac{\pd{W}_{\rm disl}}{\pd T_{ijk}},\qquad H_{ijk}=-H_{ikj}.
\end{align} 
The translational gauge field momentum tensor
$H_{ijk}$ is a very special hyperstress tensor.
It is the response quantity to the dislocation density tensor 
and has only nine independent components. 
It is convenient to perform an irreducible decomposition of the torsion,
from which we can construct the most general
linear and isotropic constitutive law between the dislocation density and the 
hyperstress in the following way~\cite{Lazar00,Lazar02a}
\begin{align}
\label{const_iso}
H_{ijk}=\sum_{I=1}^{3}a_{I}\,^{(I)}T_{ijk}.
\end{align}
Here $a_1$, $a_2$ and $a_3$ are nonnegative material coefficients due to 
the positive semi-definiteness if the dislocation core energy $W_{\rm disl}\ge 0$. They have the
dimension of a force.
The irreducible decomposition of the torsion under the rotation group $SO(3)$,
with the numbers of independent components $9=5\oplus 3\oplus 1$,
is given by
\begin{align}
T_{ijk}=\,^{(1)}T_{ijk}+\,^{(2)}T_{ijk}+\,^{(3)}T_{ijk}.
\end{align}
The tensor, the trace and the axial tensor pieces are defined
by (see also~\cite{Hehl95,Lazar02a})
\begin{alignat}{2}
\label{tentor}
^{(1)}T_{ijk}&=T_{ijk}-\,^{(2)}T_{ijk}-\,^{(3)}T_{ijk}
&&\qquad \text{(tentor)}\\
\label{trator}
^{(2)}T_{ijk}&:=\frac{1}{2}\left(\delta_{ij}T_{llk}
                         +\delta_{ik}T_{l jl}\right)
&&\qquad\text{(trator)}\\
\label{axitor}
^{(3)}T_{ijk}&:=\frac{1}{3}\left(T_{ijk}+T_{jki}+T_{kij}\right)
&&\qquad\text{(axitor).}
\end{alignat}
The hyperstress tensor may be written as
\begin{align}
\label{moment1}
H_{ijk}=c_1\, T_{ijk}
+c_2\,\big(T_{jki}+T_{kij}\big)
+c_3\,\big(\delta_{ij}T_{llk}+\delta_{ik}T_{ljl}\big)
\end{align}
with the abbreviations
\begin{align}
c_1:=\frac{1}{3}\big(2a_1+a_3\big),\qquad
c_2:=\frac{1}{3}\big(a_3-a_1\big),\qquad
c_3:=\frac{1}{2}\big(a_2-a_1\big).
\end{align}
The inverse relations are
\begin{align}
\label{EC-a}
a_1=c_1-c_2\ge 0,\qquad
a_2=c_1-c_2+2c_3\ge 0,\qquad
a_3=2c_2+c_1\ge 0.
\end{align}
Since $\beta_{ij}$ and $T_{ijk}$ are uncoupled from each other, 
the conditions of positive semi-definiteness,  
$W=W_{\text{el}}+W_{\text{disl}}\ge 0$, 
can be studied separately: $W_{\text{el}}\ge 0$ and $W_{\text{disl}}\ge 0$.
Thus, the characteristic constants of the material have to satisfy the 
conditions~(\ref{PSD1}) and (\ref{EC-a}).

With the relations $H_{im}=\frac{1}{2}\epsilon_{jkm} H_{ijk}$ and
$T_{ijk}=\epsilon_{jkn}\alpha_{in}$, we obtain
from Eq.~(\ref{const_iso}) the irreducible
decomposition
\begin{align}
\label{moment3}
H_{ij}=a_1 \Big(\underbrace{\alpha_{(ij)}-\frac{1}{3}\,
  \delta_{ij}\alpha_{kk}}_{\text{tentor:}\ ^{(1)}\alpha_{ij}}\Big)
+a_2\,\underbrace{ \alpha_{[ij]}}_{\text{trator:} \ ^{(2)}\alpha_{ij}}
+\, a_3\, \underbrace{\frac{1}{3}\,\delta_{ij}\alpha_{kk}}_{\text{axitor:} \ ^{(3)}\alpha_{ij}} .
\end{align}
Thus, the axitor represents a set of three perpendicular forests of screw
dislocations of equal strength. 
The trator describes forests of edge dislocations:
$\alpha_{xy}=-\alpha_{yx}$, $\alpha_{xz}=-\alpha_{zx}$ and  
$\alpha_{yz}=-\alpha_{zy}$. 
The tentor is the deviator of the dislocation density tensor and it
represents forests of edge dislocations:
$\alpha_{xy}=\alpha_{yx}$, $\alpha_{xz}=\alpha_{zx}$ and  
$\alpha_{yz}=\alpha_{zy}$ and screw dislocations:
$^{(1)}\alpha_{xx}=\alpha_{xx}-\frac{1}{3}(\alpha_{xx}+\alpha_{yy}+\alpha_{zz})$,
$^{(1)}\alpha_{yy}=\alpha_{yy}-\frac{1}{3}(\alpha_{xx}+\alpha_{yy}+\alpha_{zz})$,
$^{(1)}\alpha_{zz}=\alpha_{zz}-\frac{1}{3}(\alpha_{xx}+\alpha_{yy}+\alpha_{zz})$
with $^{(1)}\alpha_{xx}+ ^{(1)}\alpha_{yy} + ^{(1)}\alpha_{zz}=0$.
A similar constitutive relation was given earlier
by~\citet{HK65,Kroener69,KV92,PK01}.
Alternatively, the dislocation core energy~(\ref{W-core})
may be rewritten in the form
\begin{align}
\label{W-core2}
 W_{\rm disl}=
\frac{1}{2}\, \alpha_{ij} H_{ij}.
\end{align}
We want to note that a similar hyperstress tensor like~(\ref{moment3}) in terms of the dislocation density tensor and with three material coefficients was derived by~\citet{Claus70} in
a special case of micromorphic elasticity.
Therefore, the static gauge theory of dislocations is related to a micromorphic
theory and not to a Cosserat theory if we identify the gauge field $\phi_{ij}$
with the micro-distortion tensor of micromorphic 
elasticity~\citep{Mindlin64,Eringen99}. 
In general, a hyperstress tensor possesses 27 independent components and is the
response quantity to the gradient of the micro-distortion tensor.
The isotropic constitutive relation of such hyperstress tensor has 11
material parameters (see, e.g., \citet{Mindlin64}).

Since the tensor $H_{i[jk]}$ has only nine independent components like 
a couple stress tensor,
the following question arises:
In which way is the tensor $H_{ijk}$ related to the couple stress tensor?
The contortion tensor is defined in terms of the torsion tensor according 
to~\citep{Bilby56,Noll68,Kleinert}
\begin{align}
\label{Kont}
K_{ijk}=-\frac{1}{2}\big(T_{ijk}+T_{jki}-T_{kij}\big),\qquad K_{ijk}=-K_{jik}.
\end{align}
A moment stress tensor possesses nine independent components and for an
isotropic material the constitutive relations has just three material parameters.
The contortion tensor contains nine independent rotational degrees of freedom.
Obviously, $K_{ijk}$ is antisymmetric in $i$, $j$, whereas $T_{ijk}$ is so
in $j$, $k$.
With the identity $K_{lk}=\frac{1}{2}\epsilon_{ijl} K_{ijk}$, 
we recover the so-called Nye curvature tensor~\citep{Nye}
(see also~\citep{KS59,Kroener60,deWit70,Kroener81})\footnote{We want to mention that~\citet{Nye} derived originally the Eq.~(\ref{Nye}) if the elastic strain is zero (see also discussion in~\citet{Li07}). Here we have used the differential geometrical 
definition~(\ref{Kont}) of the contortion in order to derive~(\ref{Nye}). For non-zero elastic strain Eq.~(\ref{Nye}) was derived by~\citet{deWit70}.} 
\begin{align}
\label{Nye}
K_{ij}=\alpha_{ji}-\frac{1}{2}\, \delta_{ij}\alpha_{ll},\qquad
\alpha_{ij}=K_{ji}-\delta_{ij} K_{ll}.
\end{align}
The moment stress (or spin moment stress tensor) 
is the response of the crystal to the contortion 
\begin{align}
\label{moment-1}
\tau_{ijk}=\frac{\pd W}{\pd K_{ijk}},\qquad
\tau_{ijk}=-\tau_{jik}.
\end{align}
Thus, contortion produces moment stress and vise versa.
We can rewrite the tensor $H_{ijk}$ in terms of the moment stress tensor $\tau_{ijk}$
as follows
\begin{align}
\label{}
H_{ijk}&=-\tau_{ijk}+\tau_{jki}-\tau_{kij}\\
H_{[ij]k}&=-\tau_{ijk}.
\end{align}
Therefore, the tensor $H_{ijk}$ is given in terms of moment stresses similar like
the torsion in terms of the contortion.
Such a tensor $H_{i[jk]}$ is called pseudomoment stress tensor 
or sometimes spin energy potential tensor~\citep{HK65b,Hehl66,Hehl76,Hehl95}.
In addition, we have
\begin{align}
\label{}
H_{ij}=\tau_{ji}-\frac{1}{2}\,\delta_{ij} \tau_{ll},\qquad
\tau_{ij}=H_{ji}-\delta_{ij} H_{ll},
\end{align}
where $\tau_{ij}=\pd W/\pd K_{ij}$.
We conclude, torsion produces pseudomoment stress and vise versa.
Thus, pseudomoment stress is the specific response to dislocations.
In the following we will use the expression pseudomoment stress tensor for $H_{ijk}$.

It is known that nontrivial traction boundary problems 
in the variational formulation of the gauge theory of defects
can be formulated by means of a so-called
null Lagrangian~\cite{Edelen88,Edelen96,Edelen89}. 
When the null Lagrangian is added to the Lagrangian of elasticity, it does not change the `classical'
Euler-Lagrange equation (force equilibrium in elasticity theory) because the 
associated Euler-Lagrange equation, $\sigma^0_{ij,j}=0$, is 
identically satisfied.
After minimal replacement, $ u_{i,j}\rightarrow \beta_{ij}$, the null
Lagrangian 
\begin{align}
\label{L-null}
W_{\rm bg}=\pd_j\big(\sigma^0_{ij} u_i\big)
          =\sigma^0_{ij,j} u_i
            +\sigma^0_{ij} u_{i,j}
          \longrightarrow\sigma^0_{ij} \beta_{ij},
\end{align}
gives rise to a background stress tensor $\sigma^0_{ij}$
which can be considered as the nucleation field in the gauge theory of 
defects~\cite{Edelen96}.

The Euler-Lagrange equations of $W=W_{\text{disl}}+W_{\text{el}}-W_{\text{bg}}$ 
are given by
\begin{align}
\frac{\delta W}{\delta\beta_{ij}}\equiv
\frac{\pd W}{\pd \beta_{ij}}- 
\pd_k \frac{\pd W}{\pd\beta_{ij,k}}=0
\end{align}
which can be expressed with Eqs.~(\ref{ST}) and (\ref{moment}) as
\begin{align}
\label{ME}
H_{ijk,k} + \sigma_{ij}&=\sigma^0_{ij}& &\text{(pseudomoment equilibrium)}.
\end{align}
Thus, the null Lagrangian gives a contribution to the pseudomoment equilibrium
condition. 
Because the pseudomoment stress $H_{ijk}$ is skewsymmetric in the indices $j$ and $k$, the force
equilibrium follows from Eq.~(\ref{ME}) as a continuity equation
\begin{align}
\label{FE}
\sigma_{ij,j}&= 0& &\text{(force equilibrium)}.
\end{align}
We want to note that a similar equation like~(\ref{ME}) has been earlier obtained 
by~\citet{TS90,Kroener93} from a variational principle in nonlinear dislocation theory.
We may rewrite Eq.~(\ref{ME}) as follows
\begin{align}
\label{ME-SF}
\epsilon_{kjn}H_{in,k} = \sigma_{ij}-\sigma^0_{ij}
\end{align}
in order to see that the pseudomoment stress tensor $H_{in}$ is 
a kind of a stress function tensor of first order~\citep{Kroener58}.
Eq.~(\ref{ME}) may be decomposed into the symmetric and antisymmetric parts
\begin{align}
\label{ME-sy}
H_{(ij)k,k} + \sigma_{(ij)}&=\sigma^0_{(ij)}& &
\\
\label{ME-asy}
H_{[ij]k,k} + \sigma_{[ij]}&=\sigma^0_{[ij]}& &\text{(moment equilibrium)} .
\end{align}
Hence, Eq.~(\ref{ME-asy}) has an appearance close to Cosserat's moment of
momentum equation and $H_{[ij]k}$ has the meaning of a moment stress tensor.
The divergence of the moment stress tensor gives rise to asymmetric force stresses.
The existence of moment stresses $H_{[ij]k}$ demands
asymmetric force stresses $\sigma_{[ij]}$.
However, due to Eq.~(\ref{ME-sy}) 
the equilibrium equations of dislocation theory deviate
distinctly from those of the Cosserat theory (see also~\citet{Kroener93}).
Using Eqs.~(\ref{Tor}) and (\ref{moment1}), we can rewrite the
moment equilibrium~(\ref{ME}) in the following form
\begin{align}
\label{ME2}
&c_1(\beta_{ik,jk}- \beta_{ij,kk}) +
c_2(\beta_{ji,kk}- \beta_{jk,ki}+ \beta_{kj,ki}- \beta_{ki,kj})
\nonumber\\
&\hspace{4cm}
+c_3\big[\delta_{ij}(\beta_{lk,kl}- \beta_{ll,kk})+
\beta_{ll,ji}- \beta_{lj,li}\big] +\sigma_{ij}=\sigma^0_{ij}.
\end{align}
With the inverse constitutive relation for $\beta_{ij}$
\begin{align}
\label{CR-B}
\beta_{ij}= \frac{\gamma + \mu}{4\mu\gamma}\,\sigma_{ij} +
\frac{\gamma - \mu}{4\mu\gamma}\,\sigma_{ji}
 - \frac{\nu}{2\mu (1 + \nu)}\delta_{ij}\, \sigma_{kk}
\end{align}
where the Poisson ration $\nu$ is expressed
in terms of the Lam{\'e} coefficients
\begin{align}
\label{}
\nu=\frac{\lambda}{2\,(\lambda + \mu)},\qquad
\lambda=\frac{2\mu\nu}{1-2\nu}
\end{align}
we are able to rewrite completely the pseudomoment equilibrium condition~(\ref{ME2})
in terms of the force stress tensor according to
\begin{align}
\label{ME-S1}
 & \Big[(c_1 - c_2 + 2c_3)\,\frac{2\gamma\nu}{1 + \nu} - 2c_3\gamma\Big]
(\delta_{ij}\,\sigma_{ll,kk} -\sigma_{ll,ij})
+ 2c_3\gamma\,\delta_{ij}\,\sigma_{kl,kl} \nonumber
\\
&+\big[c_1(\gamma + \mu) - c_2(\gamma-\mu)\big](\sigma_{ik,jk}-\sigma_{ij,kk})
+ \big[c_1(\gamma - \mu) - c_2(\gamma+\mu)\big](\sigma_{ki,kj}-\sigma_{ji,kk})
\nonumber
\\&
+ \big[2c_2\mu -c_3(\gamma + \mu)\big]\,\sigma_{kj,ki}
- \big[2c_2\mu +c_3(\gamma - \mu)\big]\,\sigma_{jk,ki}
+4\mu\gamma\,\sigma_{ij}=4\mu\gamma \,\sigma^0_{ij}.
\end{align}
If we use the force equilibrium condition $\sigma_{ij,j}=0$,
Eq.~(\ref{ME-S1}) simplifies to
\begin{align}
\label{ME-S2}
 & \Big[(c_1 - c_2 + 2c_3)\,\frac{2\gamma\nu}{1 + \nu} - 2c_3\gamma\Big]
(\delta_{ij}\,\sigma_{ll,kk} -\sigma_{ll,ij})
-\big[c_1(\gamma + \mu) - c_2(\gamma-\mu)\big]\sigma_{ij,kk}
\\
&
+ \big[c_1(\gamma - \mu) - c_2(\gamma+\mu)\big](\sigma_{ki,kj}-\sigma_{ji,kk})
+ \big[2c_2\mu -c_3(\gamma + \mu)\big]\,\sigma_{kj,ki}
+4\mu\gamma\,\sigma_{ij}=4\mu\gamma \,\sigma^0_{ij}
\nonumber.
\end{align}
Equation~(\ref{ME-S2}) is the master equation, being the central equation
for the force stress tensor in this gauge theory of
dislocations.
In the next two sections,
solutions for screw and edge dislocations will be given.

\subsection{Comparison to other gauge theories of dislocations}
At this point we want to discuss the relation to other translational gauge theories of
dislocations which differ in the form of the constitutive relations of the
force stress and pseudomoment stress.
The differences between the gauge theories of dislocations are mainly the 
choice of the constitutive relations of the force and pseudomoment stresses.
Edelen~\citep{Edelen83,Edelen88,Edelen96} used a very simple constitutive
relation for the pseudomoment stress tensor.
The Edelen choice is $c_1=2s$, $c_2=0$ and $c_3=0$. 
Of course, this is not a general isotropic constitutive law.
In addition he assumed
symmetric force stresses.
Also \citet{VS88} used this choice for the constitutive relations.
\citet{Malyshev00} and \citet{Lazar02a,Lazar02b} 
discussed and used the so-called Einstein choice in three dimensions.
It is given by: $a_2=-a_1$ and $a_3=-a_1/2$. Thus, $c_1=a_1/2$, $c_2=-a_1/2$
and $c_3=-a_1$.
In the Einstein choice the pseudomoment stress tensor is symmetric in the first two
indices $H_{ijk}=H_{(ij)k}$ and is given in terms of `gradients' of the
elastic strain $\beta_{(ij)}$.
In this choice, $H_{(ij)k,k}$ reduces to the Einstein tensor. 
Another choice $c_1=a_1/2$, $c_2=-a_1/2$ and $c_3=\nu/(1-\nu) a_1$
was used by \citet{Lazar03} to calculate symmetric force stresses of screw and
edge dislocations.
It can be seen in Eq.~(\ref{EC-a}) that the Einstein-choice and also 
the latter choice 
violate the condition of nonnegative energy $W_{\rm disl}\ge 0$.
The material coefficients $a_1$, $a_2$ and $a_3$ have to be nonnegative.
But in this paper we  use the most general linear constitutive
relations~(\ref{stress}) and (\ref{moment1}) of an isotropic material with six
material parameters.
In Table~\ref{tab1} we have listed the choice of the constitutive relations.
\begin{table}[htb]
\begin{center}
\begin{tabular}{c|rrrr}
  & Edelen \cite{Edelen81}& Einstein choice \cite{Malyshev00,Lazar02a} 
  & \citet{Lazar03} & present paper\\ \hline 
$\mu$ & $\mu$ & $\mu$ & $\mu$ & $\mu$ \\ 
$\nu$ & $\nu$ & $\nu$ & $\nu$ & $\nu$ \\ 
$\gamma$ & 0 & 0 & 0 & $\gamma$ \\ \hline
$a_1$ & $2s$  & $a_1$ & $a_1$ & $a_1$ \\ 
$a_2$ & $2s$  & $-a_1$ & $\frac{1+\nu }{1-\nu} a_1$  & $a_2$\\ 
$a_3$ & $2s$  & $-\frac{a_1}{2}$ & $-\frac{a_1}{2}$  & $a_3$
\end{tabular}
\end{center}
\caption{This table lists the material parameters $a_I$, $\mu$, $\nu$ and
  $\gamma$ for 
the translational gauge theory of dislocations.}
\label{tab1}
\end{table}

\subsection{The Green tensor}
In this subsection, we want to derive the three-dimensional 
Green tensor of the master
equation~(\ref{ME-S2}). First we set $\sigma_{ij}^0=L_{ij}\delta(x)\delta(y)\delta(z)$.
The fundamental solution of Eq.~(\ref{ME-S2}) is defined by the equation
\begin{align}
\label{ME-G}
 & \Big[(c_1 - c_2 + 2c_3)\,\frac{2\gamma\nu}{1 + \nu} - 2c_3\gamma\Big]
(\delta_{ij}\,\sigma_{ll,kk} -\sigma_{ll,ij})
-\big[c_1(\gamma + \mu) - c_2(\gamma-\mu)\big]\sigma_{ij,kk}
\\
&
+ \big[c_1(\gamma - \mu) - c_2(\gamma+\mu)\big](\sigma_{ki,kj}-\sigma_{ji,kk})
+ \big[2c_2\mu -c_3(\gamma + \mu)\big]\,\sigma_{kj,ki}\nonumber\\
&\hspace{8cm}
+4\mu\gamma\,\sigma_{ij}=4\mu\gamma \,L_{ij}\, \delta(x)\delta(y)\delta(z)
\nonumber.
\end{align}
We use the notation for the three-dimensional Fourier transform \citep{GC,Wl} 
\begin{align}
\widetilde f(\qq)&\equiv {\cal{F}}_{(3)}\big[f(\rr)\big]=
\int_{-\infty}^{\infty} f(\rr)\, \e^{+\ii \qq\cdot\rr} \d\rr,\qquad
f(\rr)\equiv {\cal{F}}_{(3)}^{-1}\big[\widetilde f(\qq)\big]=
\frac{1}{(2\pi)^3}\int_{-\infty}^{\infty} \widetilde f(\qq)\, \e^{-\ii \qq\cdot\rr} \d\qq .
\end{align}
Applying the Fourier transform, it follows that
\begin{align}
\label{ME-G-FT}
 & \Big[(c_1 - c_2 + 2c_3)\,\frac{2\gamma\nu}{1 + \nu} - 2c_3\gamma\Big]
(\delta_{ij}\,q^2 \widetilde{\sigma}_{ll} -q_i q_j\widetilde{\sigma}_{ll})
-\big[c_1(\gamma + \mu) - c_2(\gamma-\mu)\big]q^2\widetilde{\sigma}_{ij}
\\
&
+ \big[c_1(\gamma - \mu) - c_2(\gamma+\mu)\big](q_k q_j \widetilde{\sigma}_{ki}-q^2\widetilde{\sigma}_{ji})
+ \big[2c_2\mu -c_3(\gamma + \mu)\big]\,q_k q_i \widetilde{\sigma}_{kj}\nonumber\\
&\hspace{8cm}
-4\mu\gamma\,\widetilde{\sigma}_{ij}=-4\mu\gamma \,L_{ij}
\nonumber.
\end{align}
The force equilibrium~(\ref{FE}) reads in the $\qq$-space: $\widetilde{\sigma}_{ij} q_j=0$.
Symmetrization, antisymmetrization, taking the trace 
and the inner product by $\qq$ from left of Eq.~(\ref{ME-G-FT})
we deduce
\begin{align}
\label{ME-G1-FT}
 & [1+\ell_1^2 q^2 ]\widetilde{\sigma}_{(ij)}
-\frac{1}{2}\,(\ell_1^2-\ell_2^2)[\delta_{ij}q^2 -q_i q_j]\widetilde{\sigma}_{ll}
-(\ell_1^2-\ell_3^2)q_k\widetilde{\sigma}_{k(i} q_{j)} =L_{(ij)}\\
\label{ME-G2-FT}
 & [1+\ell_4^2 q^2 ]\widetilde{\sigma}_{[ij]}
+(\ell_4^2-\ell_3^2)q_k \widetilde{\sigma}_{k[i}q_{j]}=L_{[ij]}\\
\label{ME-G3-FT}
& [1+\ell_2^2 q^2 ]\widetilde{\sigma}_{ll}=L_{ll}\\
\label{ME-G4-FT}
& [1+\ell_3^2 q^2] q_i\widetilde{\sigma}_{ij}=q_i L_{ij}
\end{align}
where the characteristic length scales are defined by
\begin{align}
\label{L1}
&\ell_1^2=\frac{c_1-c_2}{2\mu}=\frac{a_1}{2\mu}\\
\label{L2}
&\ell_2^2=\frac{(1-\nu)(c_1-c_2+2c_3)}{2\mu(1+\nu)}=\frac{(1-\nu)a_2}{2\mu(1+\nu)}\\
\label{L3}
&\ell_3^2=\frac{(\mu+\gamma)(c_1-c_2+c_3)}{4\mu\gamma}=\frac{(\mu+\gamma)(a_1+a_2)}{8\mu\gamma}\\
\label{L4}
&\ell_4^2=\frac{c_1+c_2}{2\gamma}=\frac{a_1+2a_3}{6\gamma}.
\end{align}
The lengths $\ell_1$, $\ell_2$ and $\ell_3$ fulfill the 
following relation
\begin{align}
\ell_3^2=\frac{\mu+\gamma}{4\gamma}\Big(\ell_1^2+\frac{1+\nu}{1-\nu}\, \ell_2^2\Big).
\end{align}
Now some remarks are in order.
In the static gauge theory of dislocations we have found four characteristic length scales.
These length scales may describe size effects produced by dislocations.
They are important in the plastic zone namely the dislocation core.
All four length scales depend on the six material parameters of an isotropic
material. 
Thus, they describe the material property on which the influence of the
pseudomoment 
stresses depends strongly. 
If the ratio of the smallest dimension of the body to the length scales is 
large, the effects of pseudomoment stresses are negligible.
However, when this ratio is not large, pseudomoment stresses may produce effects
of appreciable magnitude. This is the case in the dislocation core region.
It is interesting to
note that the internal length $\ell_1$ depends on $\mu$ and 
has a similar form as the internal
length in the couple stress theory~\citep{MT62,Mindlin63} 
and in gradient elasticity~\citep{Mindlin64}.
It can be seen that $\ell_2$ depends on the Poisson number $\nu$. This length $\ell_2$
is the characteristic length of dilatation.
Such a length appears also in gradient elasticity~\citep{Mindlin64}.
Because the characteristic lengths $\ell_3$ and $\ell_4$ depend on $\gamma$ they
look like the two characteristic
lengths of micropolar elasticity namely the characteristic lengths 
for bending and for torsion (see e.g.~\citep{Nowacki86}).
Thus, it depends on the physical problem which  length scale is of importance.
It is clear that for $\gamma\rightarrow\infty$ just the bending length, which is the
characteristic length in the theory of couple stresses,  survives
\begin{align}
\label{L3-0}
&
\lim_{\gamma\rightarrow\infty}\ell_3^2=\frac{c_1-c_2+c_3}{4\mu}=\frac{a_1+a_2}{8\mu},
\qquad
\lim_{\gamma\rightarrow\infty}\ell_4^2=0.
\end{align}
On the other hand, if $\gamma=0$, then $\ell_3\rightarrow \infty$ and $\ell_4\rightarrow \infty$. 
For the Einstein choice $a_2=-a_1$ and $a_3=-a_1/2$ the lengths convert to
\begin{align}
\label{L1-HE}
\ell_1^2=\frac{c_1}{\mu}=\frac{a_1}{2\mu},\qquad
\ell_2^2=-\frac{(1-\nu)c_1}{\mu(1+\nu)}
=-\frac{(1-\nu)a_1}{2\mu(1+\nu)},\qquad
\ell_3^2=0,\qquad
\ell_4^2=0.
\end{align}
Thus, only two lengths survive. 
These length scales $\ell_1^2$ and $\ell_2^2$ agree with 
${\cal{M}}^{-2}$ and $-{\cal{N}}^{-2}$ used by \citet{Malyshev00}.
In addition, four length scales survive in the Edelen choice together with
asymmetric force stresses
\begin{align}
\label{L1-Ed}
\ell_1^2=\frac{s}{\mu},\qquad
\ell_2^2=\frac{(1-\nu)s}{\mu(1+\nu)},\qquad
\ell_3^2=\frac{(\mu+\gamma)s}{2\mu\gamma},\qquad
\ell_4^2=\frac{s}{\gamma}.
\end{align}
The length scales $\ell_1^2$ and $\ell_2^2$ coincide with $M^{-2}$
and $N^{-2}$ introduced by \citet{Edelen81,Edelen82}.
If $\gamma=0$, then $\ell_3\rightarrow \infty$ and $\ell_4\rightarrow \infty$. 

From Eqs.~(\ref{ME-G1-FT})--(\ref{ME-G4-FT}) we obtain
\begin{align}
\label{ME-G4-FT2}
& \widetilde{\sigma}_{ll}=\frac{L_{ll}}{1+\ell_2^2 q^2 }\\
\label{ME-G5-FT2}
& q_i\widetilde{\sigma}_{ij}=\frac{q_i L_{ij}}{1+\ell_3^2 q^2 }
\end{align}
and
\begin{align}
\label{ME-G2-FT2}
 & \widetilde{\sigma}_{(ij)}
=\frac{L_{(ij)}}{1+\ell_1^2 q^2 }
+\frac{1}{2}\,(\ell_1^2-\ell_2^2)
\frac{[\delta_{ij}q^2 -q_j q_j]L_{ll}}{(1+\ell_1^2 q^2 )(1+\ell_2^2 q^2 )}
+(\ell_1^2-\ell_3^2)
\frac{q_k L_{k(i} q_{j)}}{(1+\ell_1^2 q^2 )(1+\ell_3^2 q^2 )} \\
\label{ME-G3-FT2}
 &\widetilde{\sigma}_{[ij]}
=\frac{L_{[ij]}}{1+\ell_4^2 q^2 }
-(\ell_4^2-\ell_3^2)
\frac{q_k L_{k[i} q_{j]}}{(1+\ell_4^2 q^2 )(1+\ell_3^2 q^2 )} .
\end{align}

Using the inverse Fourier transformed function~\citep{GC,Wl} 
\begin{align}
\label{FT-e}
{\cal {F}}^{-1}_{(3)}\Big[\frac{1}{q^2+\frac{1}{\ell^2}}\Big]&=\frac{1}{4\pi r}\,\e^{-r/\ell},\qquad r=\sqrt{x^2+y^2+z^2}
\end{align}
we find
\begin{align}
\label{}
\sigma_{(ij)}&=
\frac{L_{(ij)}}{4\pi r\ell_1^2}\, \e^{-r/\ell_1}
-\frac{L_{kk}}{8\pi}\, [\delta_{ij}\Delta-\pd_i\pd_j]
\Big[\frac{1}{r}\big(\e^{-r/\ell_1}-\e^{-r/\ell_2}\big)\Big]
-\frac{1}{4\pi}\, L_{k(i}\pd_{j)}\pd_k
\Big[\frac{1}{r}\big(\e^{-r/\ell_1}-\e^{-r/\ell_3}\big)\Big]\\
\label{}
\sigma_{[ij]}&=
\frac{L_{[ij]}}{4\pi r\ell_4^2}\, \e^{-r/\ell_4}
+\frac{1}{4\pi}\, L_{k[i}\pd_{j]}\pd_k\Big[\frac{1}{r}\big(\e^{-r/\ell_4}-\e^{-r/\ell_3}\big)\Big].
\end{align}
With $\sigma_{ij}=\sigma_{(ij)}+\sigma_{[ij]}$, the fundamental
solution can be written in the following form
\begin{align}
\sigma_{ij}=G_{ijkl} L_{kl}
\end{align}
where $G_{ijkl}$ is the Green tensor of Eq.~(\ref{ME-G}).
So, the Green tensor reads
\begin{align}
\label{GT}
G_{ijkl}&=
\frac{1}{8\pi}\bigg\{
(\delta_{ik}\delta_{jl}+\delta_{il}\delta_{jk})
\,\frac{\e^{-r/\ell_1}}{ \ell_1^2 r}
+(\delta_{ik}\delta_{jl}-\delta_{il}\delta_{jk})
\,\frac{\e^{-r/\ell_4}}{ \ell_4^2 r}
-(\delta_{ij}\Delta-\pd_i\pd_j)\delta_{kl}
\Big[\frac{1}{r}\big(\e^{-r/\ell_1}-\e^{-r/\ell_2}\big)\Big]\nonumber\\
&\quad
-(\delta_{jl}\pd_i+\delta_{il}\pd_j)\pd_k
\Big[\frac{1}{r}\big(\e^{-r/\ell_1}-\e^{-r/\ell_3}\big)\Big]
-(\delta_{jl}\pd_i-\delta_{il}\pd_j)\pd_k
\Big[\frac{1}{r}\big(\e^{-r/\ell_4}-\e^{-r/\ell_3}\big)\Big]
\bigg\}.
\end{align}
Using the convolution theorem, we obtain for the solution of Eq.~(\ref{ME-S2})
\begin{align}
\sigma_{ij}=G_{ijkl}*\sigma_{kl}^0 .
\end{align}
With the help of this equation three-dimensional problems can be solved
e.g. dislocation loops.

\section{Screw dislocation}
\setcounter{equation}{0}
\label{screw}
We consider now a straight screw dislocation. We choose the dislocation line
and the Burgers vector in $z$-direction: $b_x=b_y=0$, $b_z=b$.
First of all, a screw dislocation corresponds to the anti-plane strain problem.
Thus, we only have the non-vanishing components of the elastic distortion
tensor $\beta_{zx}$ and $\beta_{zy}$.
From the constitutive equation~(\ref{stress}) we obtain
the relations valid in anti-plane strain:
\begin{align}
\label{APS}
\sigma_{zx}=\frac{\mu-\gamma}{\mu+\gamma}\, \sigma_{xz},\qquad
\sigma_{zy}=\frac{\mu-\gamma}{\mu+\gamma}\, \sigma_{yz}.
\end{align}
In the case of a screw dislocation we want to solve Eq.~(\ref{ME2}) directly.
From the force equilibrium condition~(\ref{FE}), the following condition
follows
\begin{align}
\label{BC}
\beta_{zx,x} + \beta_{zy,y}=0.
\end{align}
It looks like a gauge condition, but it is nothing else than a consequence
of the force equilibrium condition~(\ref{FE}).
Using Eq.~(\ref{BC}), (\ref{ME2}) simplifies to
\begin{alignat}{2}
\label{}
&\Big[1- \frac{c_1}{\mu +\gamma}\Delta\Big]\beta_{zx} = \beta^0_{zx},
\hspace{2cm}
&&\Big[1+ \frac{c_2}{\mu -\gamma}\Delta\Big]\beta_{zx} = \beta^0_{zx},\\
&\Big[1- \frac{c_1}{\mu +\gamma}\Delta\Big]\beta_{zy} = \beta^0_{zy},
&&\Big[1+ \frac{c_2}{\mu-\gamma}\Delta\Big]\beta_{zy}=\beta^0_{zy}.
\end{alignat}
Here $\Delta$ denotes the two-dimensional Laplacian.
Because we have four equations for two components $\beta_{zx}$ and
$\beta_{zy}$, we obtain a relation between $c_2$ and $c_1$ as 
follows\footnote{If $\gamma=0$,
we obtain $c_2=-c_1$.
If $c_2=0$ like in the Edelen choice, we obtain $c_1=0$,
$\ell_1^2=0$, $\beta_{zx}=\beta^0_{zx}$ and $\beta_{zy}=\beta^0_{zy}$. 
Thus, only the classical solution of a screw dislocation 
is allowed in the Edelen choice with symmetric force stresses. 
\citet{Edelen96} used some ad hoc assumptions 
like $\beta^0_{xz}=\beta_{zx}$ to avoid this problems.
But in the classical theory it must be: $\beta^0_{xz}=0$.
Also the approach of~\citet{VS88} possesses such lacks.} 
\begin{align}
\label{Rel-screw}
c_2= - \frac{\mu - \gamma}{\mu + \gamma}\, c_1
\end{align}
or in the `irreducible' parameters
\begin{align}
\label{Rel-screw-ir}
a_3=\frac{3\gamma-\mu}{2\mu}\, a_1.
\end{align}
Relation~(\ref{Rel-screw}) is a consequence of~(\ref{APS}).
From the relation~(\ref{Rel-screw-ir}) together with the condition of 
non-negative dislocation core energy~(\ref{EC-a}) we obtain the constraint between
$\mu$ and $\gamma$:
\begin{align}
\gamma\ge \frac{\mu}{3}.
\end{align}
It is important to note that in the case of $\gamma=0$, $a_3=-a_1/2$ 
violates the condition~(\ref{EC-a}).
Therefore, the condition of nonnegative dislocation core energy 
demands asymmetric force stresses.
For the Edelen choice ($a_3=a_1$) we obtain from (\ref{Rel-screw}) and (\ref{Rel-screw-ir}):
$\mu=\gamma$. Thus, it does not make sense to use symmetric force stresses
in the Edelen model. 
Due to these reasons, the Edelen model demands asymmetric force stresses 
with $\mu=\gamma$ for a screw dislocation.
Therefore, with $\gamma=\mu$ the correct solution of a screw dislocation of the Edelen model is contained in our general solution of the screw dislocation which we calculate in the following. 

If we substitute (\ref{Rel-screw}) and (\ref{Rel-screw-ir}) in (\ref{L1}) and (\ref{L4}),
we obtain the characteristic length scale of the anti-plane strain problem
\begin{align}
\label{L1-2}
\ell_1^2=\ell_4^2= \frac{c_1}{\mu + \gamma}=\frac{a_1}{2\mu}
\end{align}
and the equations for the elastic distortion simplify to
\begin{align}
\label{HE-Bzx}
&\big[1 - \ell_1^2\Delta\big]\beta_{zx} = \beta^0_{zx},\qquad
\big[1 - \ell_1^2\Delta\big]\beta_{zy} = \beta^0_{zy}.
\end{align}
They are two-dimensional inhomogeneous Helmholtz equations.
The inhomogeneous parts are given by Eq.~(\ref{B-0}).
If we substitute (\ref{B-0}) into Eq.~(\ref{HE-Bzx}),  
the elastic distortion is easily obtained as
\begin{align}
\label{B-zx}
\beta_{zx}&= -\frac{b}{2\pi}\frac{y}{r^2}
\Big[1-\frac{r}{\ell_1}K_1\Big(\frac{r}{\ell_1}\Big)\Big],\qquad
\beta_{zy}= \frac{b}{2\pi}\frac{x}{r^2}
\Big[1-\frac{r}{\ell_1}K_1\Big(\frac{r}{\ell_1}\Big)\Big]
\end{align}
where $K_n$ is the modified Bessel
function of the second kind and $n=0,1,\ldots$ denotes the order of this function.
Substituting the elastic distortions~(\ref{B-zx}) 
into (\ref{Tor}), the torsion tensor or dislocation tensor is calculated as
\begin{align}
\label{Tor-screw}
T_{zxy}=\alpha_{zz} =\frac{b}{2\pi\ell^2_1}\,K_0\Big(\frac{r}{\ell_1}\Big).
\end{align}
The torsion~(\ref{Tor-screw}) generates the following components of the
pseudomoment stress tensor~(\ref{moment1}) according
\begin{align}
H_{xx}&=H_{xyz}=c_2\,T_{zxy} =-\frac{(\mu - \gamma)b}{2\pi}\,K_0\Big(\frac{r}{\ell_1}\Big)\\
H_{yy}&=H_{yzx}=c_2\,T_{zxy} =-\frac{(\mu - \gamma)b}{2\pi}\,K_0\Big(\frac{r}{\ell_1}\Big)\\
\label{}
H_{zz}&=H_{zxy}=c_1\,T_{zxy} = \frac{(\mu + \gamma)b}{2\pi}\,K_0\Big(\frac{r}{\ell_1}\Big).
\end{align}
Thus, this localized pseudomoment stress caused by a screw dislocation is of
torsion-type. 
It is given in terms of only one length scale $\ell_1$.
The trace of the pseudomoment stresses is
\begin{align}
H_{kk}=-\frac{(\mu - 3\gamma)b}{2\pi}\,K_0\Big(\frac{r}{\ell_1}\Big).
\end{align}

Now we want to show how we can solve the master
equation~(\ref{ME-S2}) by means of a suitable stress function
ansatz for the asymmetric force stresses of a screw dislocation.
With the stress function ansatz we fulfill the force equilibrium
condition~(\ref{FE}). Then we substitute the stress function
ansatz into the pseudomoment equilibrium condition~(\ref{ME-S2}). In the
case of a straight screw dislocation Eq.~(\ref{ME-S2}) reduces to
\begin{align}
\label{ME-Tzx}
&\big[c_1(\gamma - \mu) - c_2(\gamma+\mu)\big](\sigma_{yz,yx}-\sigma_{xz,yy})
- \big[c_1(\gamma + \mu) - c_2(\gamma-\mu)\big]\Delta\sigma_{zx}
+ 4\mu\gamma\sigma_{zx}=4\mu\gamma\sigma^0_{zx}\\
\label{ME-Tzy}
&\big[c_1(\gamma - \mu) - c_2(\gamma+\mu)\big](\sigma_{xz,xy} -\sigma_{yz,xx})
- \big[c_1(\gamma + \mu) - c_2(\gamma-\mu)\big]\Delta\sigma_{zy}
+ 4\mu\gamma\sigma_{zy}=4\mu\gamma\sigma^0_{zy}\\
\label{ME-Txz}
&\big[c_2(\gamma + \mu) - c_1(\gamma-\mu)\big]\Delta\sigma_{zx}
+ \big[c_2(\gamma - \mu) - c_1(\gamma+\mu)\big]\Delta\sigma_{xz}
\\
&\hspace{6cm}
+\big[2c_2\mu -c_3(\gamma + \mu)\big](\sigma_{xz,xx} + \sigma_{yz,xy})
+ 4\mu\gamma\sigma_{xz}= 4\mu\gamma\sigma^0_{xz}
\nonumber\\
\label{ME-Tyz}
&\big[c_2(\gamma + \mu) - c_1(\gamma-\mu)\big]\Delta\sigma_{zy}
+ \big[c_2(\gamma - \mu) - c_1(\gamma+\mu)\big]\Delta\sigma_{yz}
\\
&\hspace{6cm}
+\big[2c_2\mu -c_3(\gamma + \mu)\big](\sigma_{xz,xy} + \sigma_{yz,yy})
+ 4\mu\gamma\sigma_{yz}= 4\mu\gamma\sigma^0_{yz}.
\nonumber
\end{align}
We introduce the following stress function ansatz,
which is suitable for asymmetric force stresses in the anti-plane strain problem,
\begin{align}
\label{SFA-F}
\sigma_{ij}=
\left(\begin{array}{ccc}
0 & 0 & -\frac{\mu - \gamma}{\mu + \gamma}\,\pd_{y} F\\
0 & 0 & \ \  \frac{\mu - \gamma}{\mu + \gamma}\,\pd_{x} F \\
-\pd_{y} F &  \pd_{x} F & 0
\end{array}\right)
\end{align}
where $F$ is a modified Prandtl stress function.
Using (\ref{Rel-screw}), (\ref{L1}) and (\ref{F-0}), we have to solve the two-dimensional inhomogeneous Helmholtz
equation
\begin{align}
\label{F-dgl}
\big[1 - \ell_1^2\Delta\big]F = F^0.
\end{align}
Its solution reads
\begin{align}
\label{F}
F = \frac{(\mu + \gamma)b}{2\pi}\,\Big[\ln r + K_0\Big(\frac{r}{\ell_1}\Big)\Big].
\end{align}
With Eqs.~(\ref{SFA-F}) and (\ref{F}) the force stress is calculated as
\begin{alignat}{2}
\label{}
\sigma_{zx}&= -\frac{(\mu +\gamma)b}{2\pi}\frac{y}{r^2}\Big[1-\frac{r}{\ell_1}K_1\Big(\frac{r}{\ell_1}\Big)\Big],\qquad
&&\sigma_{xz}= -\frac{(\mu - \gamma)b}{2\pi}\frac{y}{r^2}\Big[1-\frac{r}{\ell_1}K_1\Big(\frac{r}{\ell_1}\Big)\Big],\\
\sigma_{zy}&= \frac{(\mu + \gamma)b}{2\pi}\frac{x}{r^2}\Big[1-\frac{r}{\ell_1}K_1\Big(\frac{r}{\ell_1}\Big)\Big], \qquad
\label{}
&&\sigma_{yz}= \frac{(\mu - \gamma)b}{2\pi}\frac{x}{r^2}\Big[1-\frac{r}{\ell_1}K_1\Big(\frac{r}{\ell_1}\Big)\Big].
\end{alignat}
In cylindrical coordinates the non-vanishing components of the force stress and
elastic distortion tensors read
\begin{align}
\label{Tzp}
\sigma_{z\phi}&= \frac{(\mu +\gamma)b}{2\pi}\,\frac{1}{r}
\Big[1-\frac{r}{\ell_1}K_1\Big(\frac{r}{\ell_1}\Big)\Big],\qquad
\sigma_{\phi z}= \frac{(\mu -\gamma)b}{2\pi}\,\frac{1}{r}
\Big[1-\frac{r}{\ell_1}K_1\Big(\frac{r}{\ell_1}\Big)\Big]\\
\label{Bzp}
\beta_{z\phi}&= \frac{b}{2\pi}\,\frac{1}{r}
\Big[1 - \frac{r}{\ell_1}K_1\Big(\frac{r}{\ell_1}\Big)\Big].
\end{align}
\begin{figure}[t]\unitlength1cm
\vspace*{-0.5cm}
\centerline{
\begin{picture}(8,6)
\put(0.0,0.2){\epsfig{file=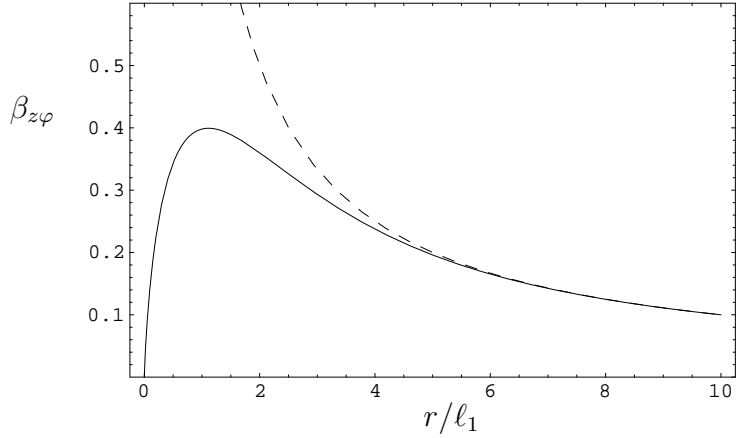,width=9cm}}
\put(4.5,-0.1){$r/\ell_1$}
\put(-1.0,4.0){$\beta_{z\varphi}$}
\end{picture}
}
\caption{The component~$\beta_{z\varphi}$ is given in units of $ b/[2\pi\ell_1]$ .  
The dashed curve represents the classical component.}
\label{fig:Dist}
\end{figure} 
The force stress and elastic distortion fields, calculated in the gauge theory,
do not possess singularities. In fact, they have at $r\simeq 1.114 \ell_1$ a maximum 
of: $\sigma_{z\phi}\simeq 0.399 (\mu+\gamma)b/[2\pi \ell_1]$, 
$\sigma_{\phi z}\simeq 0.399 (\mu-\gamma)b/[2\pi \ell_1]$, 
$\beta_{z\phi}\simeq 0.399 b/[2\pi \ell_1]$ (see figure~\ref{fig:Dist}).
We want to note that if $\gamma=0$, we recover the force stress and 
pseudomoment stress, the elastic distortion and the dislocation tensor earlier calculated by 
\citet{Malyshev00} and \citet{Lazar02a,Lazar02b} 
in the gauge theory of dislocations with the so-called Einstein choice.
The modified Burgers vector is given by
\begin{align}
\label{Burger-screw}
b(r)=\oint^{2\pi}_0\beta_{z\phi}\,r\,\d\phi
= \int^{2\pi}_0\int^r_0 \alpha_{zz}(r')\,r'\,\d
r'\,\d\phi= b\Big[1-\frac{r}{\ell_1}K_1\Big(\frac{r}{\ell_1}\Big)\Big].
\end{align}
This modified Burgers vector differs appreciably from the constant
value $b$ in the core region from $r=0$ up to $r\simeq 6\ell_1$. Thus, it is
suggestive to take $r_{\text{c}}\simeq 6 \ell_1$ as the dislocation core radius.
Outside this core region, the Burgers vector reaches its constant value.
Accordingly, the gauge theoretical solutions approaches
the classical one outside the dislocation core region.
Due to the cylindrical symmetry of a straight screw dislocation 
around the dislocation line the physical situation requires only one length 
scale $\ell_1$.

\begin{figure}[t]\unitlength1cm
\vspace*{-0.5cm}
\centerline{
\begin{picture}(8,6)
\put(0.0,0.2){\epsfig{file=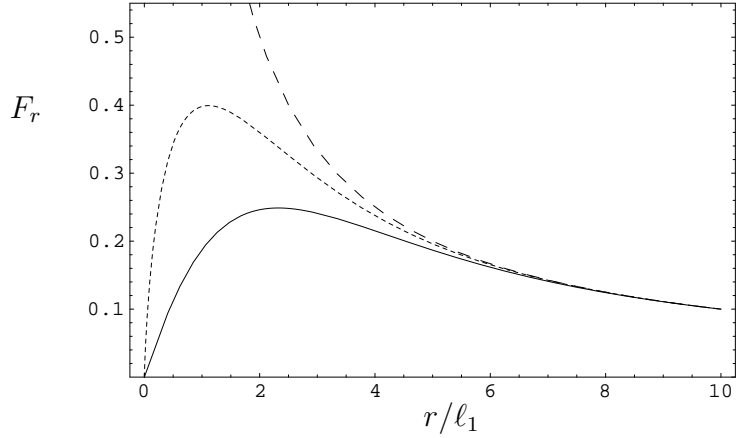,width=9cm}}
\put(4.5,-0.1){$r/\ell_1$}
\put(-1.0,4.0){$F_{r}$}
\end{picture}
}
\caption{The component~$F_{r}$ is given in units of $ \mu bb'/[2\pi\ell_1]$ ($\gamma=0$).  
The small dashed curve represents the nonlocal result and the 
dashed curve represents the classical component.}
\label{fig:F}
\end{figure}

The force between dislocations is the so-called Peach-Koehler force~\cite{PK,Maugin93,Kirchner}.
In the gauge theory of dislocations
the gauge invariant formulation of the 
Peach-Koehler force has been derived by~\citet{AL07}.
\citet{Edelen83} and \citet{Edelen88} just derived an expression which is not
gauge invariant because they used the canonical Eshelby stress tensor instead of
the gauge invariant one.
It is given by the convolution of the stress and the
dislocation density tensor
\begin{align}
\label{PKF}
F_k=-\sigma_{li}*T_{lik}=\epsilon_{kij}\sigma_{li}*\alpha_{lj}.
\end{align}
It is obvious that (\ref{PKF}) is similar in the form as the classical expression
of the Peach-Koehler force.
But, in (\ref{PKF}) the stress and the dislocation density tensor 
are the gauge theoretical ones.
For a screw dislocation it reads
\begin{align}
F_x=\sigma_{zy}*\alpha_{zz},\qquad
F_y=-\sigma_{zx}*\alpha_{zz}.
\end{align}
By the convolution theorem, it is in the Fourier space
\begin{align}
\widetilde{F}_x(\qq)={\cal {F}}_{(2)} \big[F_x]=\widetilde{\sigma}_{zy}(\qq)\widetilde{\alpha}_{zz}(\qq),\qquad
\widetilde{F}_y(\qq)={\cal {F}}_{(2)} \big[F_y]=-\widetilde{\sigma}_{zx}(\qq)\widetilde{\alpha}_{zz}(\qq).
\end{align}
We use the following notation for the two-dimensional Fourier transform \citep{GC,Wl} 
\begin{align}
\widetilde f(\qq)\equiv {\cal{F}}_{(2)}\big[f(\rr)\big]=
\int_{-\infty}^{\infty} f(\rr)\, \e^{+\ii \qq\cdot\rr} \d\rr, 
\qquad
f(\rr)\equiv {\cal{F}}_{(2)}^{-1}\big[\widetilde f(\qq)\big]=
\frac{1}{(2\pi)^2}\int_{-\infty}^{\infty} \widetilde f(\qq)\, \e^{-\ii \qq\cdot\rr} \d\qq .
\end{align}
The Fourier transformed functions are
\begin{align}
\label{FT-ln}
{\cal {F}}_{(2)}\big[\ln r\big]=-\frac{2\pi}{q^2},
\qquad
{\cal {F}}_{(2)}\Big[ K_0\Big(\frac{r}{\ell_1}\Big)\Big]=\frac{2\pi}{q^2+\frac{1}{\ell_1}}
\end{align}
and we obtain
\begin{align}
\widetilde{\sigma}_{zy}(\qq)&=(\mu+\gamma)b\,  
\frac{\ii q_x}{\ell_1^2 q^2\big(q^2+\frac{1}{\ell_1}\big)},\qquad
\widetilde{\sigma}_{zx}(\qq)=-(\mu+\gamma)b\,  
\frac{\ii q_y}{\ell_1^2 q^2\big(q^2+\frac{1}{\ell_1}\big)}\\
\widetilde{\alpha}_{zz}(\qq)&=b' 
\frac{1}{\ell_1^2 \big(q^2+\frac{1}{\ell_1}\big)}.
\end{align}
The Fourier transformed Peach-Koehler force $\widetilde{F}_x$ is
\begin{align}
\widetilde{F}_{x}(\qq)&=(\mu+\gamma)b b'\,
\frac{\ii q_x}{\ell_1^4 q^2\big(q^2+\frac{1}{\ell_1}\big)^2}.
\end{align}
Using the inverse Fourier transform, the Peach-Koehler force $F_x$ is
calculated as
\begin{align}
F_{x}
&=-(\mu+\gamma)b b'\,  
\pd_x\, {\cal {F}}_{(2)}^{-1}\bigg[\frac{1}{q^2}-\frac{1}{q^2+\frac{1}{\ell_1}}
-\frac{1}{\ell_1^2 \big(q^2+\frac{1}{\ell_1}\big)^2}
\bigg]\nonumber\\
&=(\mu+\gamma)b b'\,  \pd_x \Big[\ln r+K_0\Big(\frac{r}{\ell_1}\Big)
+\frac{r}{2\ell_1} K_1\Big(\frac{r}{\ell_1}\Big)\Big]\nonumber\\
&=\frac{(\mu+\gamma)b b'}{2\pi}\, \frac{x}{r^2} \Big[1-\frac{r}{\ell_1} K_1\Big(\frac{r}{\ell_1}\Big)
-\frac{r^2}{2\ell^2_1} K_0\Big(\frac{r}{\ell_1}\Big)\Big].
\end{align}
After an analogous computation
we obtain for the component $F_y$
\begin{align}
F_y=\frac{(\mu+\gamma)b b'}{2\pi}\, \frac{y}{r^2} \Big[1-\frac{r}{\ell_1} K_1\Big(\frac{r}{\ell_1}\Big)
-\frac{r^2}{2\ell^2_1} K_0\Big(\frac{r}{\ell_1}\Big)\Big].
\end{align}
In polar coordinates, the non-vanishing component reads:
\begin{align}
\label{PKF-r}
F_r=\frac{(\mu+\gamma)b b'}{2\pi}\, \frac{1}{r} \Big[1-\frac{r}{\ell_1} K_1\Big(\frac{r}{\ell_1}\Big)
-\frac{r^2}{2\ell^2_1} K_0\Big(\frac{r}{\ell_1}\Big)\Big].
\end{align}
The Peach-Koehler force between two screw dislocations is a radial force. 
It has a maximum of: $F_r\simeq 0.2488 (\mu+\gamma)b b'/[2\pi \ell_1]$ at $r\simeq 2.324 \ell_1$ 
and is zero at $r=0$ (see figure~\ref{fig:F}).
It is interesting to note that the maximum of the Peach-Koehler 
force calculated in the framework of nonlocal elasticity is higher than our
gauge theoretical result (see \citet{Eringen83,Eringen02,Lazar05}).
If one calculates the gradient of the elastic interaction energy between two screw 
dislocations, then also the result~(\ref{PKF-r}) follows. 
But, if one uses the `total' interaction energy including the core 
contribution, one does not get the Peach-Koehler force.
Because \citet{VS96} used the `total' interaction energy, they have not calculated
the Peach-Koehler force in the gauge theory. 
Nevertheless, their result is in agreement with the nonlocal expression
calculated by \citet{Eringen83}.  

\section{Edge dislocation}
\setcounter{equation}{0}
\label{edge}
In this section we consider a straight edge dislocation. 
We choose the dislocation line in the $z$-direction and the Burgers vector reads
$b_x=b$,  $b_y=0$, $b_z=0$. 
The problem of an edge dislocation corresponds to plane strain.


\subsection{The general case}
Let us turn to the edge dislocation in the framework of the
gauge theory of dislocations. In this case the master
equation~(\ref{ME2}) reduces to
\begin{align}
\label{ME-xx}
 & \big[(c_1 - c_2 + 2c_3)\,2\gamma\nu - 2c_3\gamma(1+\nu)\big](\sigma_{xx,yy}+\sigma_{yy,yy})
-2\gamma (c_1- c_2)\Delta\sigma_{xx}
\\
&\hspace{2cm}
+ \big[(c_1-c_2)(\gamma - \mu) -c_3(\gamma + \mu)\big](\sigma_{xx,xx}+ \sigma_{yx,yx})
+4\mu\gamma\,\sigma_{xx}=4\mu\gamma \,\sigma^0_{xx}
\nonumber\\
\label{ME-yy}
 & \big[(c_1 - c_2 + 2c_3)\,2\gamma\nu - 2c_3\gamma(1+\nu)\big](\sigma_{xx,xx}+\sigma_{yy,xx})
-2\gamma (c_1- c_2)\Delta\sigma_{yy}
\\
&\hspace{2cm}
+ \big[(c_1-c_2)(\gamma - \mu) -c_3(\gamma + \mu)\big](\sigma_{xy,xy}+ \sigma_{yy,yy})
+4\mu\gamma\,\sigma_{yy}=4\mu\gamma \,\sigma^0_{yy}
\nonumber\\
\label{ME-xy}
 &
\big[c_1(\gamma - \mu) - c_2(\gamma+\mu)\big](\sigma_{xx,xy}-\sigma_{yx,xx})
+ \big[2c_2\mu -c_3(\gamma + \mu)\big](\sigma_{xy,xx}+\sigma_{yy,yx} )
\nonumber\\
&\hspace{2cm}
-\big[(c_1 - c_2 + 2c_3)\,2\gamma\nu - 2c_3\gamma(1+\nu)\big](\sigma_{xx,xy}+\sigma_{yy,xy})
\\
&\hspace{5cm}
-\big[c_1(\gamma + \mu) - c_2(\gamma-\mu)\big]\Delta \sigma_{xy}
+4\mu\gamma\,\sigma_{xy}=4\mu\gamma \,\sigma^0_{xy}
\nonumber\\
\label{ME-yx}
 &
\big[c_1(\gamma - \mu) - c_2(\gamma+\mu)\big](\sigma_{yy,yx}-\sigma_{yx,yy})
+ \big[2c_2\mu -c_3(\gamma + \mu)\big](\sigma_{xx,xy}+\sigma_{yx,yy} )
\nonumber\\
&\hspace{2cm}
-\big[(c_1 - c_2 + 2c_3)\,2\gamma\nu - 2c_3\gamma(1+\nu)\big](\sigma_{xx,xy}+\sigma_{yy,xy})
\\
&\hspace{5cm}
-\big[c_1(\gamma + \mu) - c_2(\gamma-\mu)\big]\Delta \sigma_{yx}
+4\mu\gamma\,\sigma_{yx}=4\mu\gamma \,\sigma^0_{yx}
\nonumber\\
\label{ME-zz}
&\Big[1-\frac{(1-\nu)c_3}{2\mu\nu}\Delta\Big]\sigma_{zz}=\sigma^0_{zz}.
\end{align}
We used the plane strain condition $\sigma_{zz}=\nu(\sigma_{xx}+\sigma_{yy})$ because 
we require plane strain for an edge dislocation in the gauge theory.

For the asymmetric force stress we use the stress function ansatz of
Mindlin-type with the two modified stress functions $f$ and $\Psi$:
\begin{align}
\label{SFA-edge}
\sigma_{ij}=
\left(\begin{array}{ccc}
\pd^2_{yy}f - \pd^2_{xy}\Psi  & -\pd^2_{xy}f + \pd^2_{xx}\Psi  & 0 \\\\
 -\pd^2_{xy}f - \pd^2_{yy}\Psi & \pd^2_{xx}f + \pd^2_{xy}\Psi  & 0 \\\\
 0  &  0  & \nu \Delta f
\end{array} \right).
\end{align}
This stress function ansatz~(\ref{SFA-edge}) has been also used by~\citet{LM04}
in gradient micropolar elasticity.
If we substitute~(\ref{SFA-edge}) into the coupled system
of equations~(\ref{ME-xx})--(\ref{ME-zz}), we obtain the decoupled
equations
\begin{align}
\label{ME-xx-S}
&\frac{\pd^2}{\pd{y^2}}
\Big[\Big(1 - \frac{(c_1 - c_2 +c_3)(1-\nu)}{2\mu}\Delta\Big)f - f^0\Big]
\nonumber\\
& \hspace{3cm}
- \frac{\pd^2}{\pd{x}\pd{y}}
\Big[\Big(1 - \frac{(c_1 - c_2+c_3)(\mu +\gamma)}{4\mu\gamma}\Delta\Big)\Psi
- \Psi^0\Big] =0\\
\label{ME-yy-S}
&
\frac{\pd^2}{\pd{x^2}}
\Big[\Big(1 -\frac{(c_1 - c_2 +c_3)(1-\nu)}{2\mu}\Delta\Big)f- f^0\Big]
\nonumber\\
&\hspace{3cm}
+\frac{\pd^2}{\pd{x}\pd{y}}
\Big[\Big(1 - \frac{(c_1- c_2 + c_3)(\mu + \gamma)}{4\mu\gamma}\Delta\Big)\Psi
- \Psi^0\Big] =0\\
\label{ME-xy-S}
&\frac{\pd^2}{\pd{x}\pd{y}}
\Big[\Big(1 - \frac{(c_1 - c_2 +c_3)(1-\nu)}{2\mu}\Delta\Big)f - f^0\Big]
\nonumber\\
&\hspace{3cm}
-\frac{\pd^2}{\pd{x^2}}
\Big[\Big(1 - \frac{(c_1 -c_2 + c_3)(\mu + \gamma)}{4\mu\gamma}\Delta\Big)\Psi
- \Psi^0\Big] =0\\
\label{ME-yx-S}
&\frac{\pd^2}{\pd{x}\pd{y}}
\Big[\Big(1 - \frac{(c_1 - c_2 +c_3)(1-\nu)}{2\mu}\Delta\Big)f - f^0\Big]
\nonumber\\
&\hspace{3cm}
+\frac{\pd^2}{\pd{y^2}}
\Big[\Big(1 - \frac{(c_1 -  c_2 + c_3)(\mu + \gamma)}{4\mu\gamma}\Delta\Big)\Psi
- \Psi^0\Big] =0\\
\label{ME-zz-S}
&\Delta \Big[ \Big(1 - \frac{c_3(1 - \nu)}{2\mu \nu} \Delta\Big)f - f^0\Big] =0.
\end{align}
The addition of Eqs.~(\ref{ME-xx-S}) and (\ref{ME-yy-S}) gives
\begin{align}
\label{ME-add}
\Delta \Big[\Big(1 - \frac{(c_1 - c_2 +  c_3)(1-\nu)}{2\mu}\Delta\Big)f - f^0\Big] =0.
\end{align}
For the moment equilibrium condition~(\ref{ME-asy}), we obtain
\begin{align}
\label{ME-asy-Psi}
\Delta \Big[\Big(1 - \frac{(c_1 - c_2 +  c_3)(\mu+\gamma)}{4\mu\gamma}\Delta\Big)\Psi - \Psi^0\Big] =0.
\end{align}
If we compare (\ref{ME-zz-S}) with (\ref{ME-add}), we obtain the following
constraint between $c_1$, $c_2$ and $c_3$:\footnote{This relation 
is not fulfilled in the Einstein choice and $c_1-c_2+c_3=0$.  
In the Edelen choice $c_3=0$, $c_2=0$, we obtain from (\ref{Rel-edge}) $c_1=0$,
$\ell_2=0$ and $\ell_3=0$. Thus, only $f=f^0$ and $\Psi=\Psi^0$ are allowed with the
Edelen choice 
if the plane strain condition is fulfilled. Due to these reasons, \citet{Edelen82,Edelen83,Malyshev00} used a stress function ansatz not fulfilling the plane strain condition.}
\begin{align}
\label{Rel-edge}
c_3= \frac{\nu}{1-\nu}\,(c_1 - c_2)
\end{align}
or in `irreducible' parameters
\begin{align}
\label{Rel-edge-a}
a_2=\frac{1+\nu}{1-\nu}\, a_1 .
\end{align}
This constraint is nothing but
a consequence of the relation $\sigma_{zz}=\nu(\sigma_{xx}+\sigma_{yy})$ valid in
the plane strain problem.
We want to note that this constraint (\ref{Rel-edge-a}) has been earlier 
derived by~\citet{Lazar03}.
With $-1\le\nu\le 1/2$, we get from~(\ref{Rel-edge-a}) the relation
\begin{align}
0\le a_2\le 3 a_1.
\end{align}
If we substitute (\ref{Rel-edge}) and (\ref{Rel-edge-a}) into
(\ref{L2}) and (\ref{L3}), we obtain the two characteristic length scales
of the plane strain problem
\begin{align}
\label{L2-2}
\ell^2_2&=\frac{c_1 - c_2}{2\mu}=\frac{1}{2\mu}\, a_1
 \\
\label{L3-2}
\ell^2_3&= \frac{(c_1 - c_2)(\mu +  \gamma)}{4\mu(1 - \nu)\gamma}
= \frac{\mu +  \gamma}{4\mu(1 - \nu)\gamma}\, a_1 .
\end{align}
The relation between these internal lengths is
\begin{align}
\ell^2_3=\frac{\mu+\gamma}{2(1-\nu)\gamma}\, \ell^2_2 .
\end{align}
Therefore, we obtain from Eqs.~(\ref{ME-xx-S})--(\ref{ME-zz-S})
two inhomogeneous Helmholtz equations which we have to solve
\begin{align}
\label{HE-f}
[1 - \ell^2_2\,\Delta]f &= f^0\\
\label{HE-psi}
[1 - \ell^2_3\,\Delta]\Psi &= \Psi^0
\end{align}
where the inhomogeneous parts are given by
(\ref{f0}) and (\ref{psi0}).
The solutions of Eqs.~(\ref{HE-f}) and (\ref{HE-psi}) are the following
modified `Airy' stress functions
\begin{align}
\label{f}
f&= -\frac{\mu b}{4\pi(1-\nu)}\,\pd_y \Big\{ r^2\ln r
+ 4\,\ell^2_2\Big[\ln r +K_0\Big(\frac{r}{\ell_2}\Big)\Big]\Big\} \\
\label{psi} 
\Psi&= \frac{\mu\gamma\,b}{2\pi(\mu + \gamma)}\, \pd_x
\Big\{r^2 \ln r + 4\,\ell^2_3\Big[\ln r
+K_0\Big(\frac{r}{\ell_3}\Big)\Big]\Big\}.
\end{align}

\begin{figure}[t]\unitlength1cm
\centerline{
\epsfig{figure=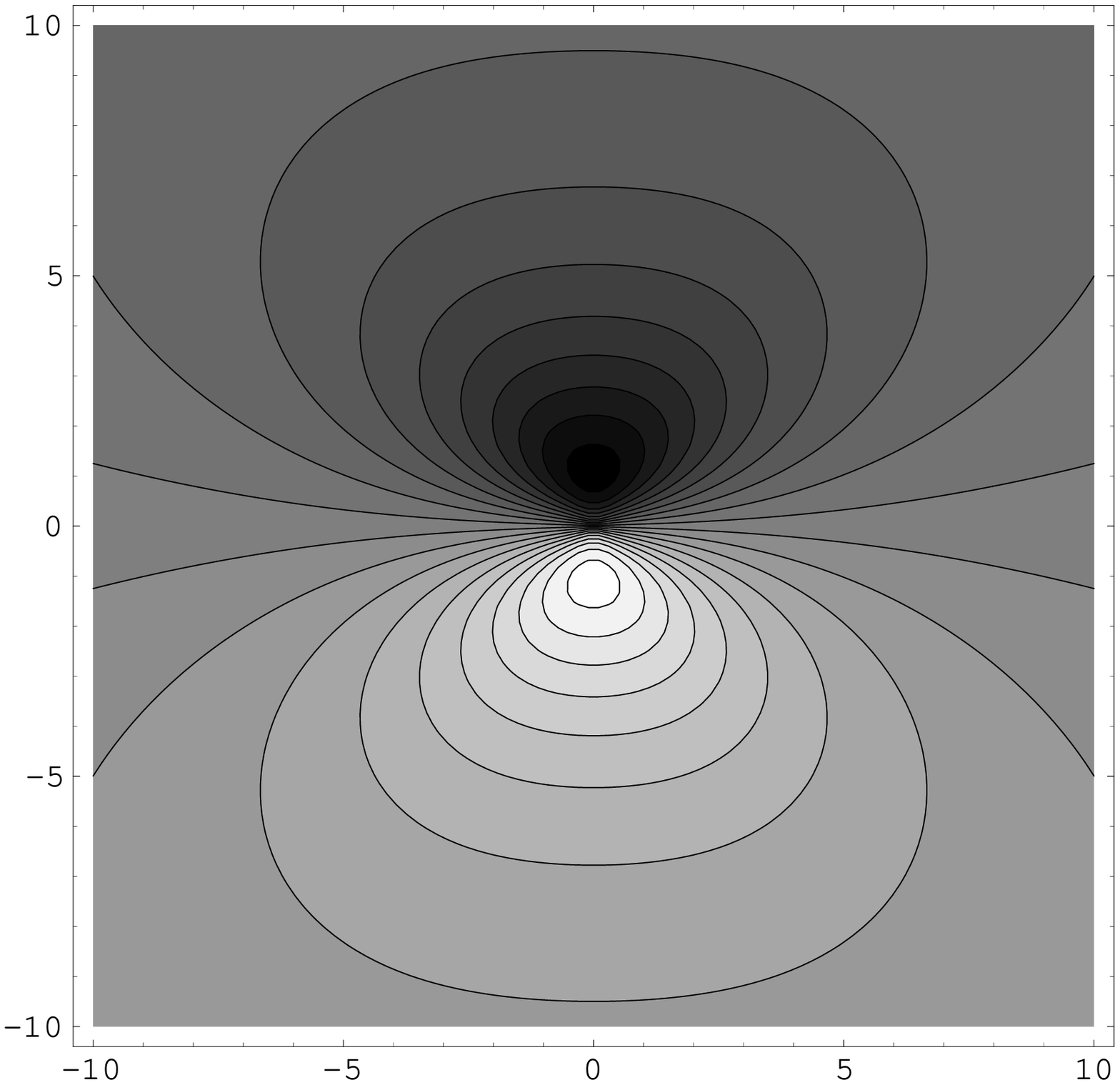,width=6.0cm}
\put(-6.7,3.0){$y/\ell_2$}
\put(-6.2,-0.3){$\text{(a)}$}
\hspace*{0.2cm}
\put(0,-0.3){$\text{(b)}$}
\epsfig{figure=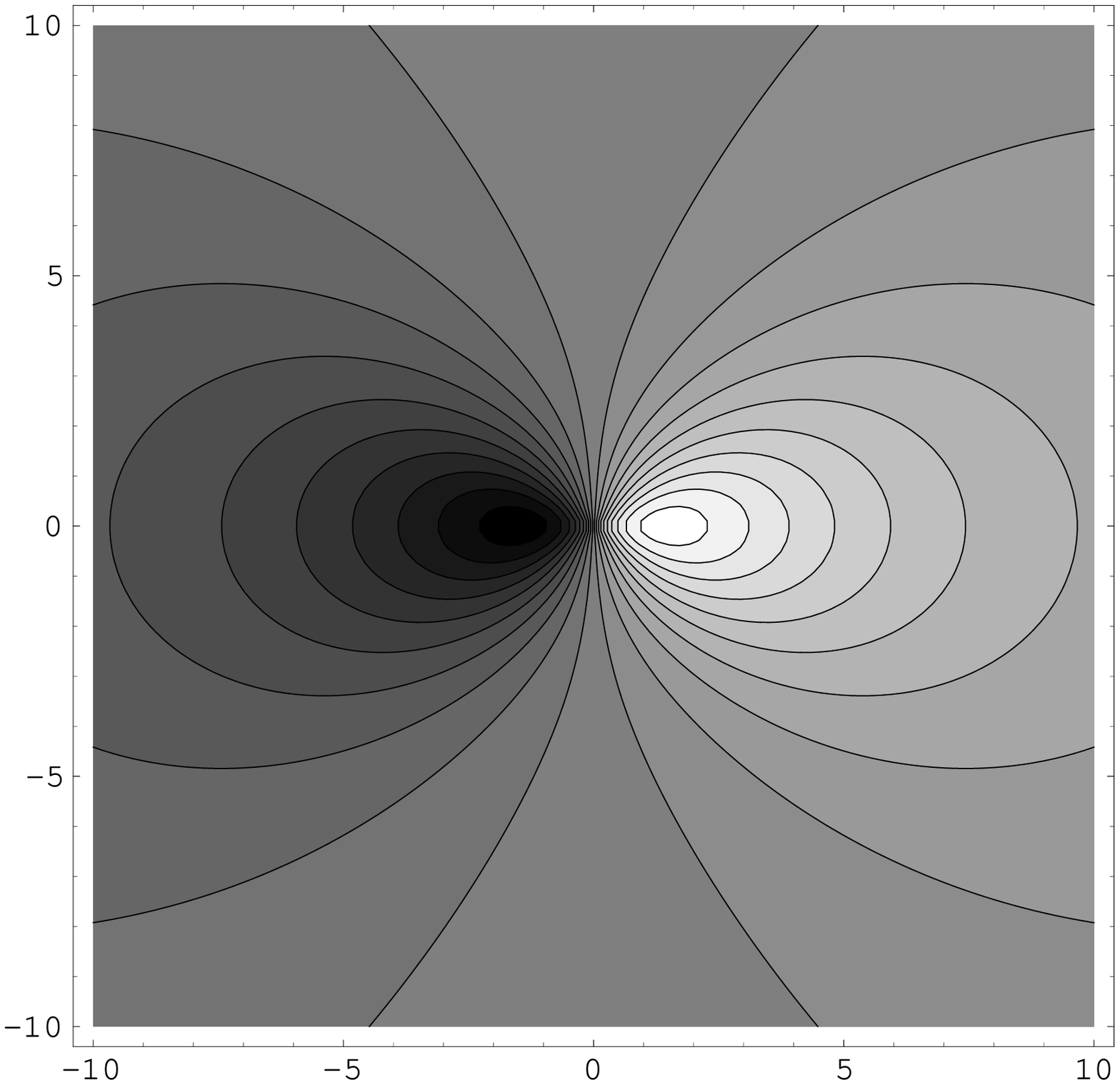,width=6.0cm}
}
\vspace*{0.2cm}
\centerline{
\epsfig{figure=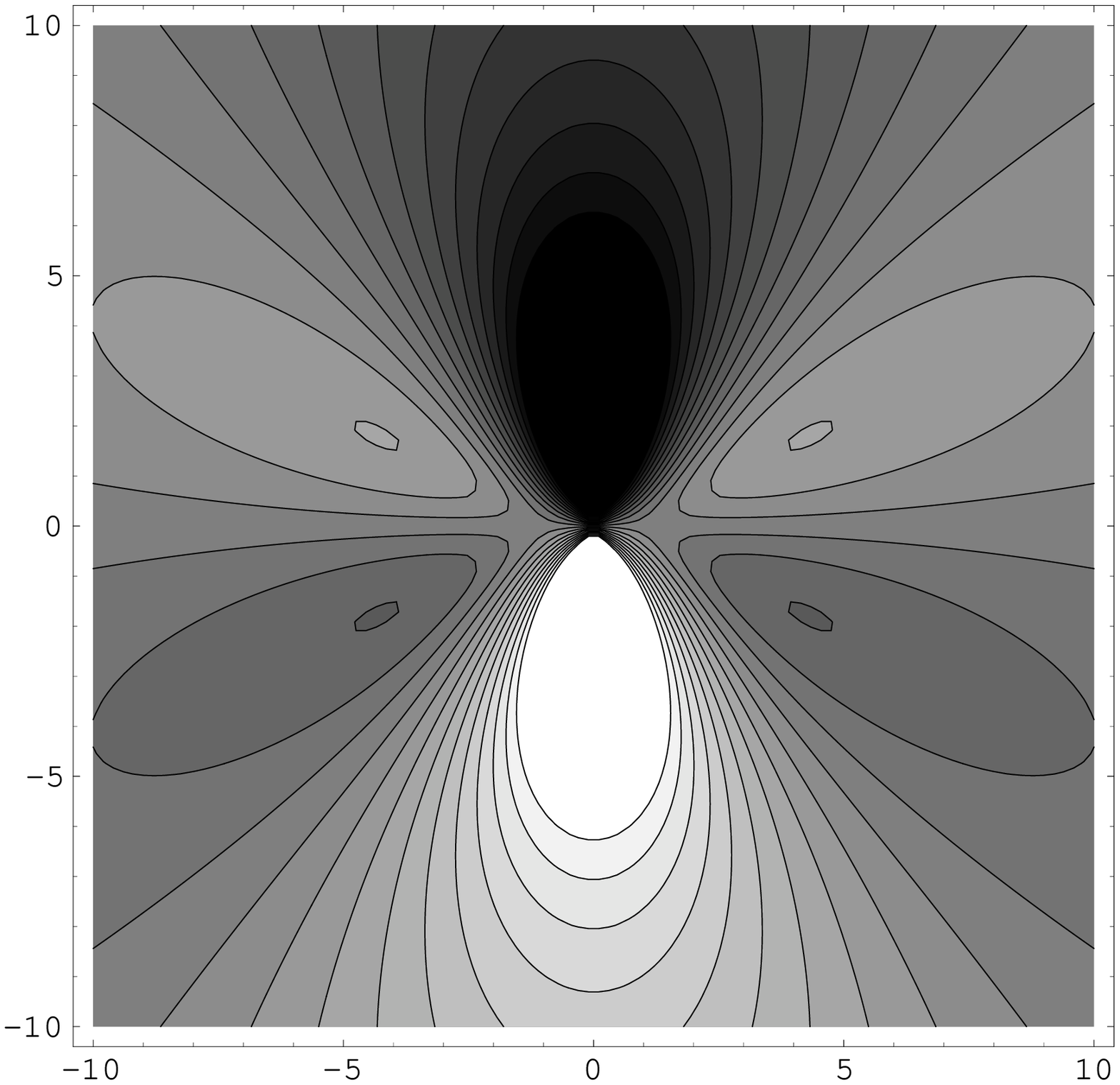,width=6.0cm}
\put(-3.0,-0.3){$x/\ell_2$}
\put(-6.7,3.0){$y/\ell_2$}
\put(-6.0,-0.3){$\text{(c)}$}
\hspace*{0.2cm}
\put(3.0,-0.3){$ x/\ell_2$}
\put(0,-0.3){$\text{(d)}$}
\epsfig{figure=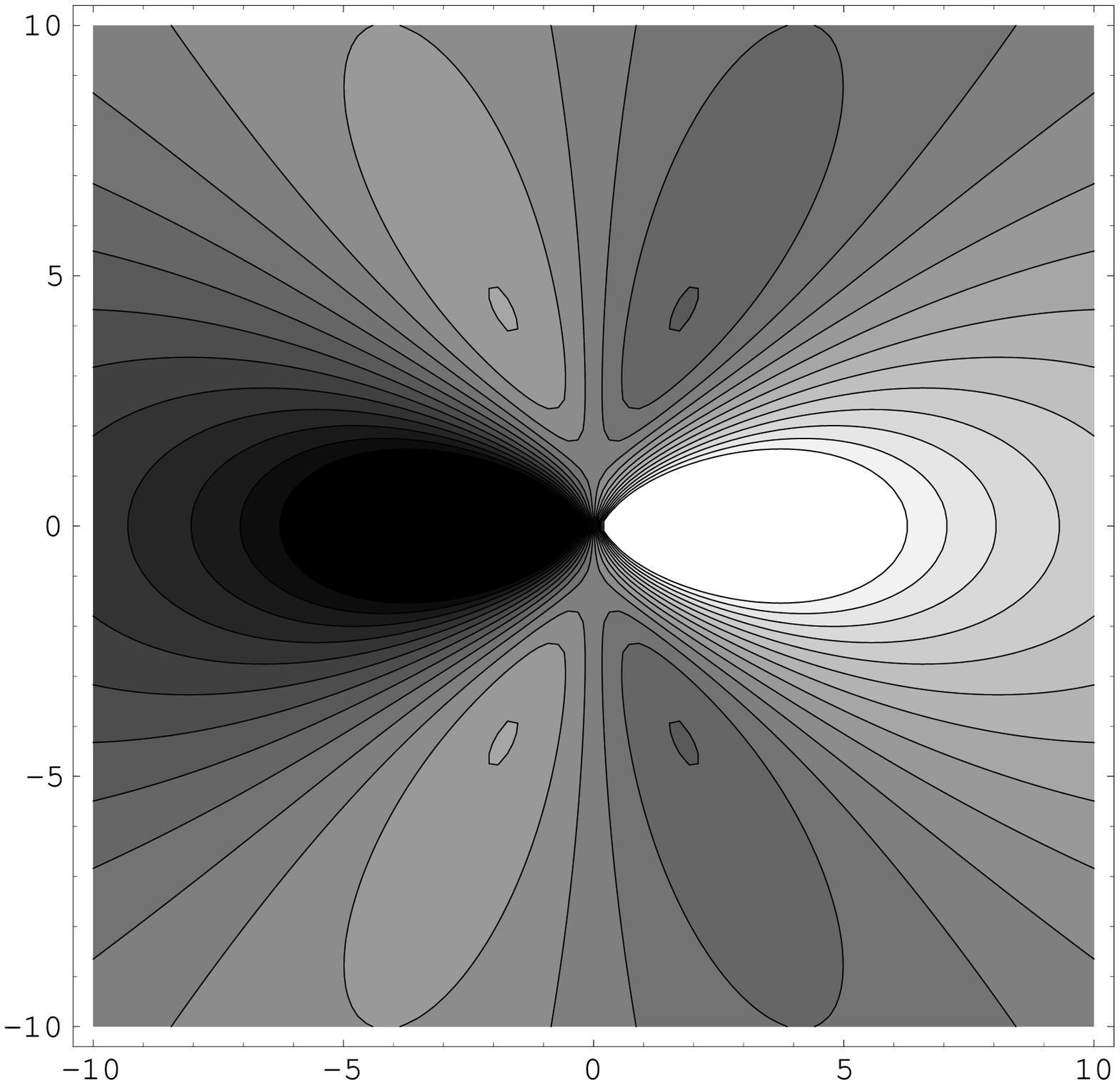,width=6.0cm}
}
\caption{Force stress contours of an edge dislocation near the dislocation line:
(a) $\sigma_{xx}$,
(b) $\sigma_{xy}$,
(c) $\sigma_{yy}$,
(d) $\sigma_{yx}$ with $\nu=0.3$ and $\gamma=\mu/2$.}
\label{fig:stress-Co}
\end{figure}

\begin{figure}[t]\unitlength1cm
\centerline{
\epsfig{figure=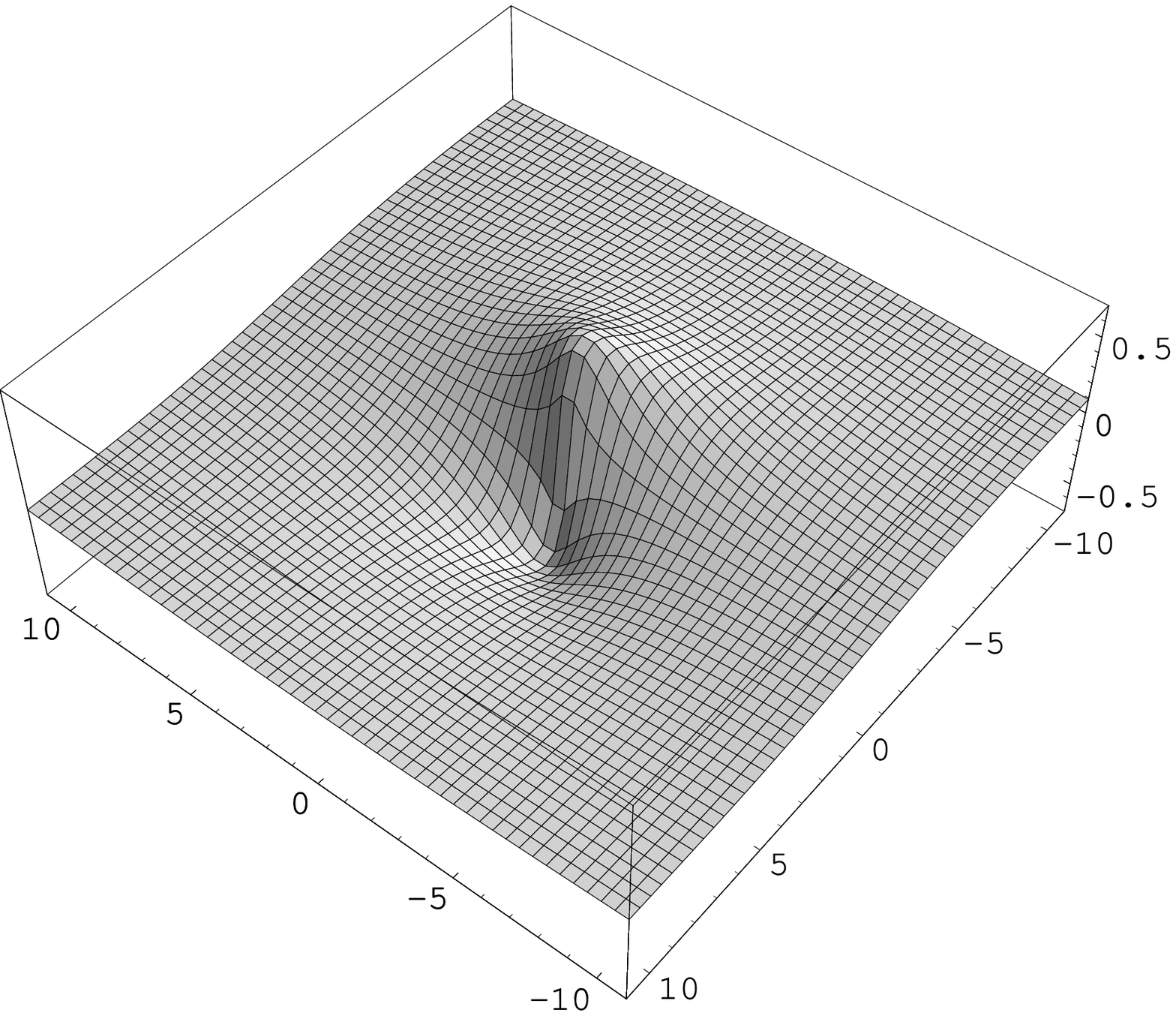,width=7.0cm}
\put(-1.5,1.0){$y/\ell_2$}
\put(-6.5,1.0){$x/\ell_2$}
\put(-6.2,-0.3){$\text{(a)}$}
\hspace*{0.4cm}
\put(0,-0.3){$\text{(b)}$}
\put(5.5,1.0){$y/\ell_2$}
\put(0.5,1.0){$x/\ell_2$}
\epsfig{figure=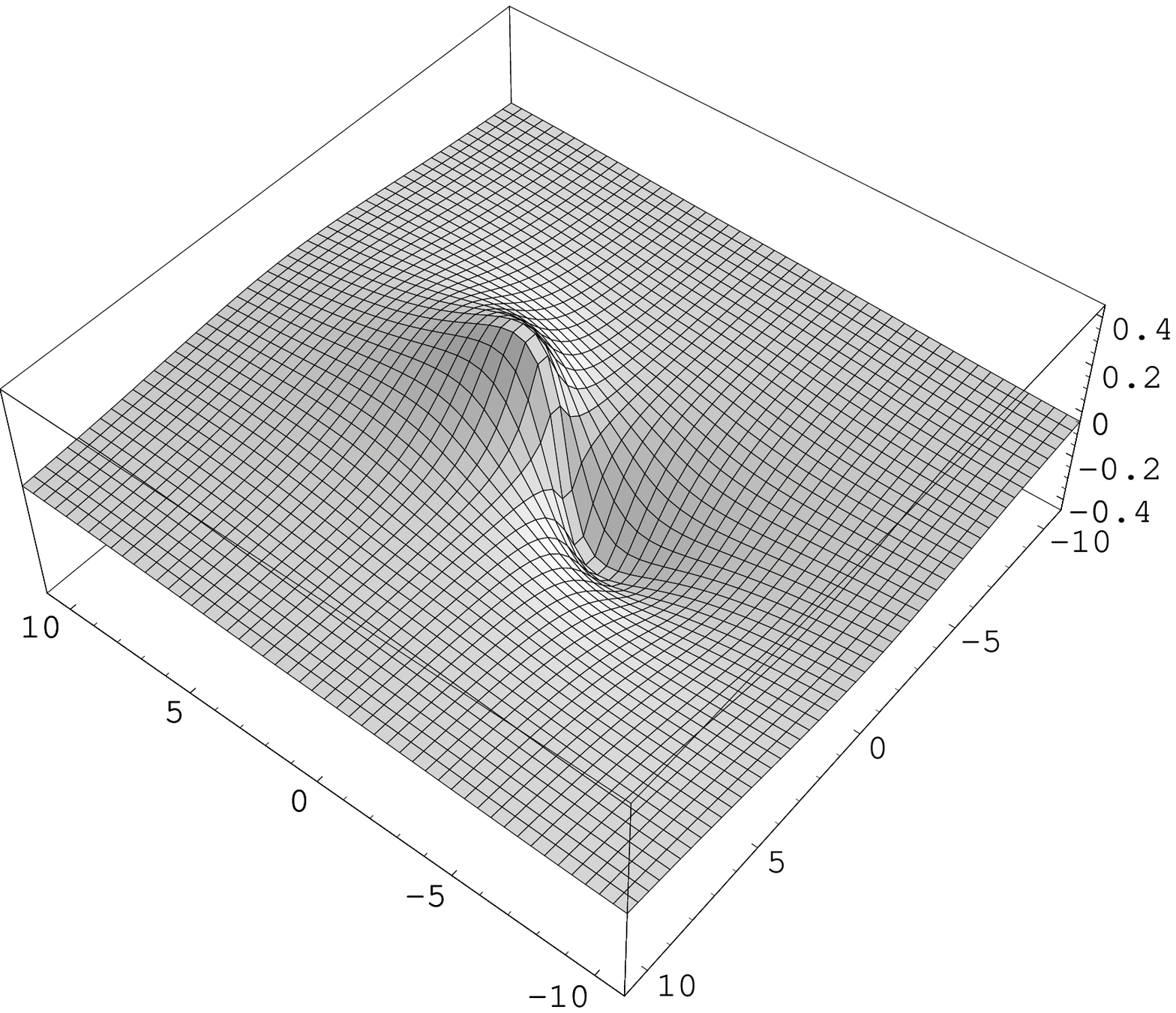,width=7.0cm}
}
\vspace*{0.2cm}
\centerline{
\epsfig{figure=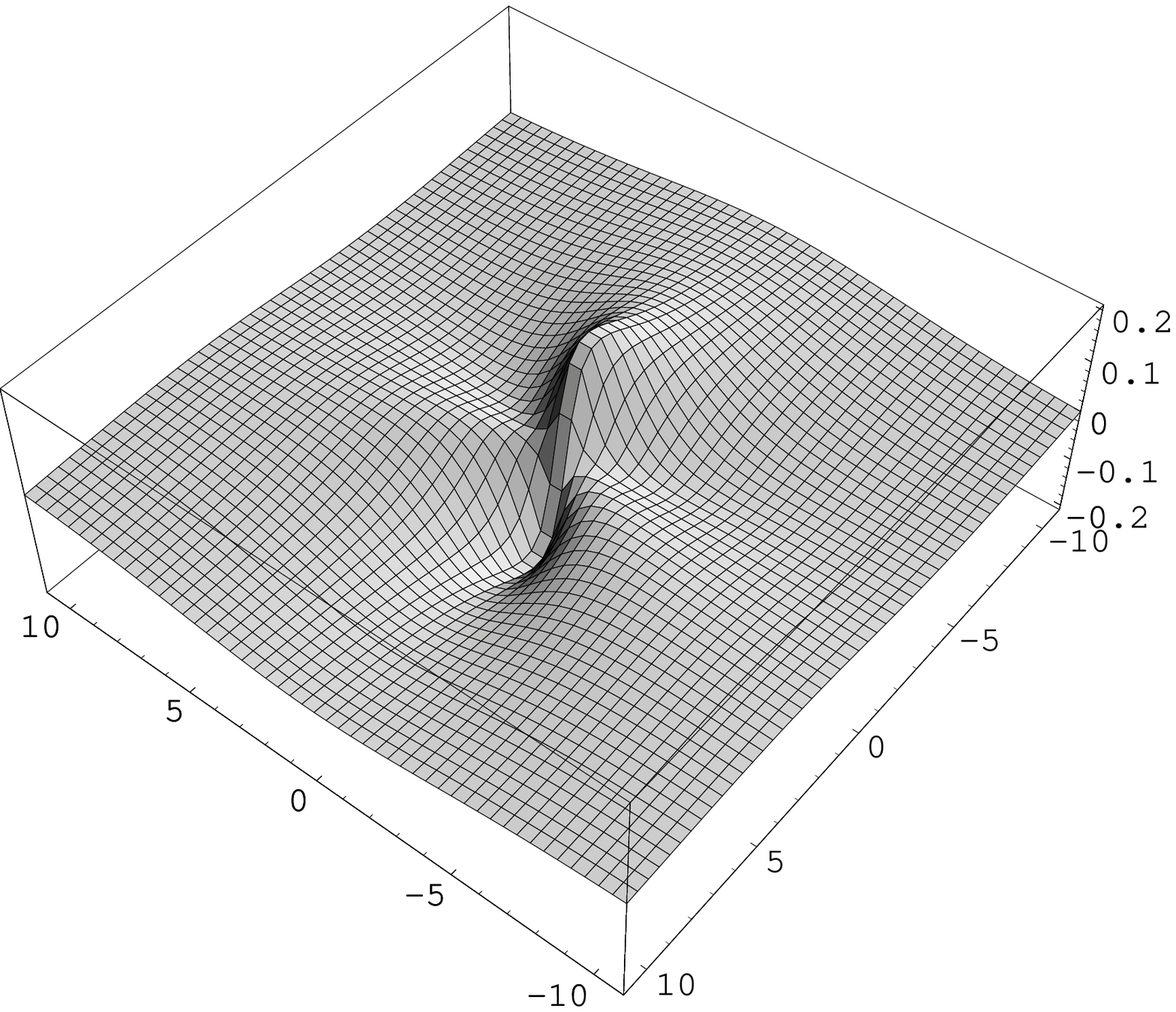,width=7.0cm}
\put(-1.5,1.0){$y/\ell_2$}
\put(-6.5,1.0){$x/\ell_2$}
\put(-6.0,-0.3){$\text{(c)}$}
\hspace*{0.4cm}
\put(5.5,1.0){$y/\ell_2$}
\put(0.5,1.0){$x/\ell_2$}
\put(0,-0.3){$\text{(d)}$}
\epsfig{figure=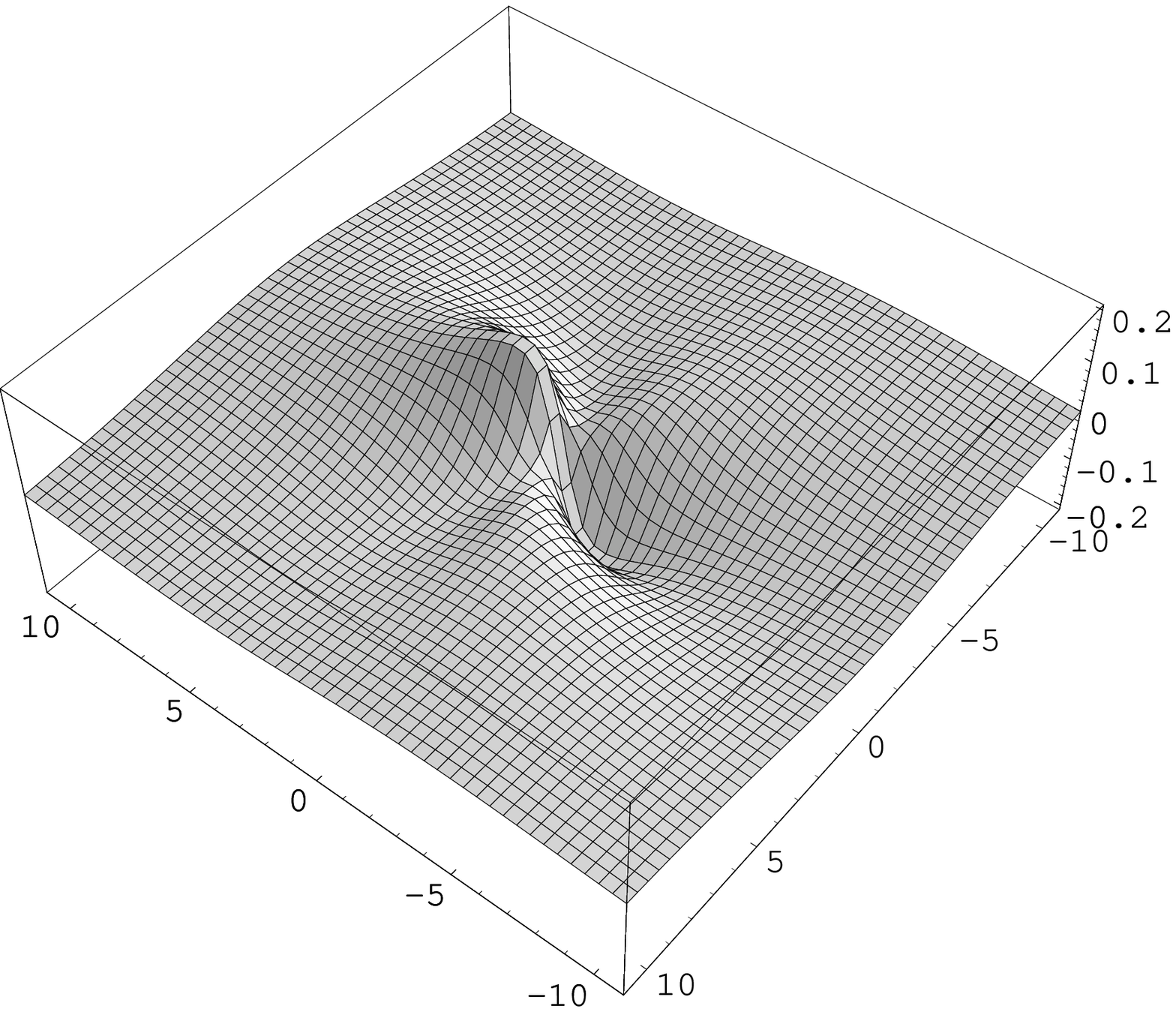,width=7.0cm}
}
\caption{Force stress of an edge dislocation:
(a) $\sigma_{xx}$,
(b) $\sigma_{xy}$,
(c) $\sigma_{yy}$,
(d) $\sigma_{yx}$
are given in units of $A$ with $\nu=0.3$ and $\gamma=\mu/2$.}
\label{fig:stress-3D}
\end{figure}

\begin{figure}[tp]\unitlength1cm
\vspace*{-1.0cm}
\centerline{
(a)\begin{picture}(8,6)
\put(0.0,0.2){\epsfig{file=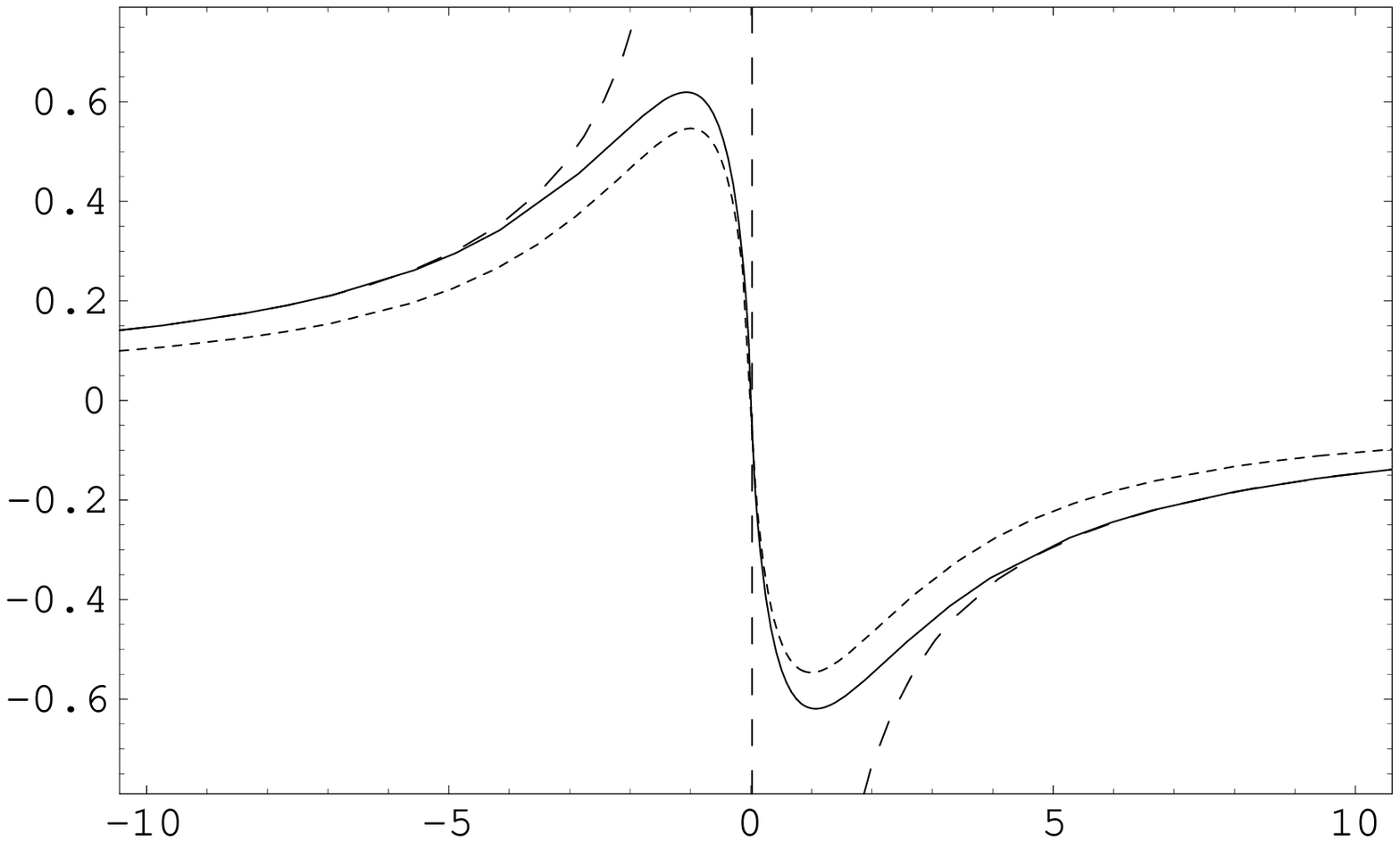,width=8cm}}
\put(4.0,0.0){$y/\ell_2$}
\put(-1.5,2.5){$\sigma_{xx}(0,y)$}
\end{picture}}
\vspace*{-1.0cm}
\centerline{
(b)\begin{picture}(8,6)
\put(0.0,0.2){\epsfig{file=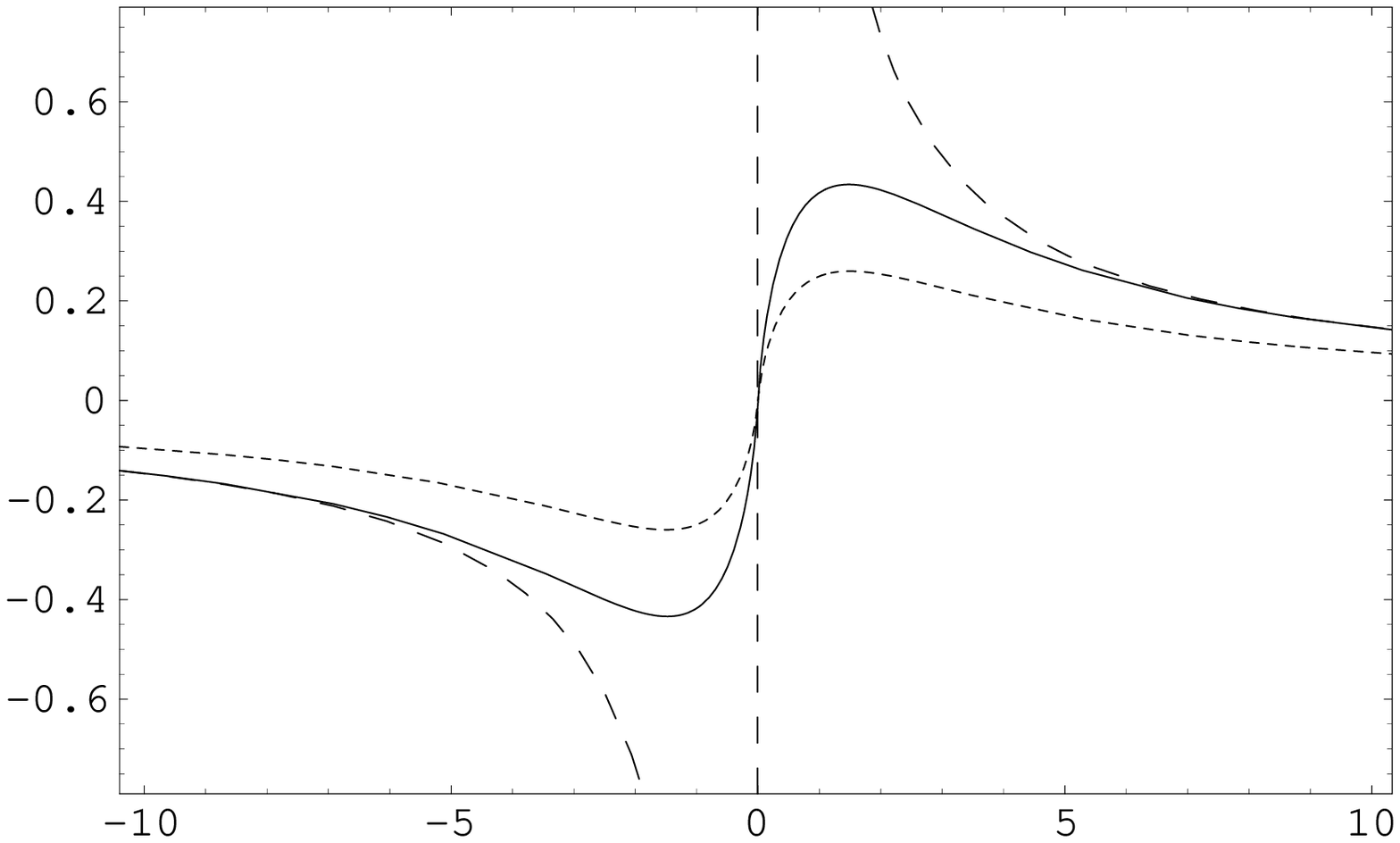,width=8cm}}
\put(4.0,0.0){$x/\ell_2$}
\put(-1.5,2.5){$\sigma_{xy}(x,0)$}
\end{picture}}
\vspace*{-1.0cm}
\centerline{
(c)
\begin{picture}(8,6)
\put(0.0,0.2){\epsfig{file=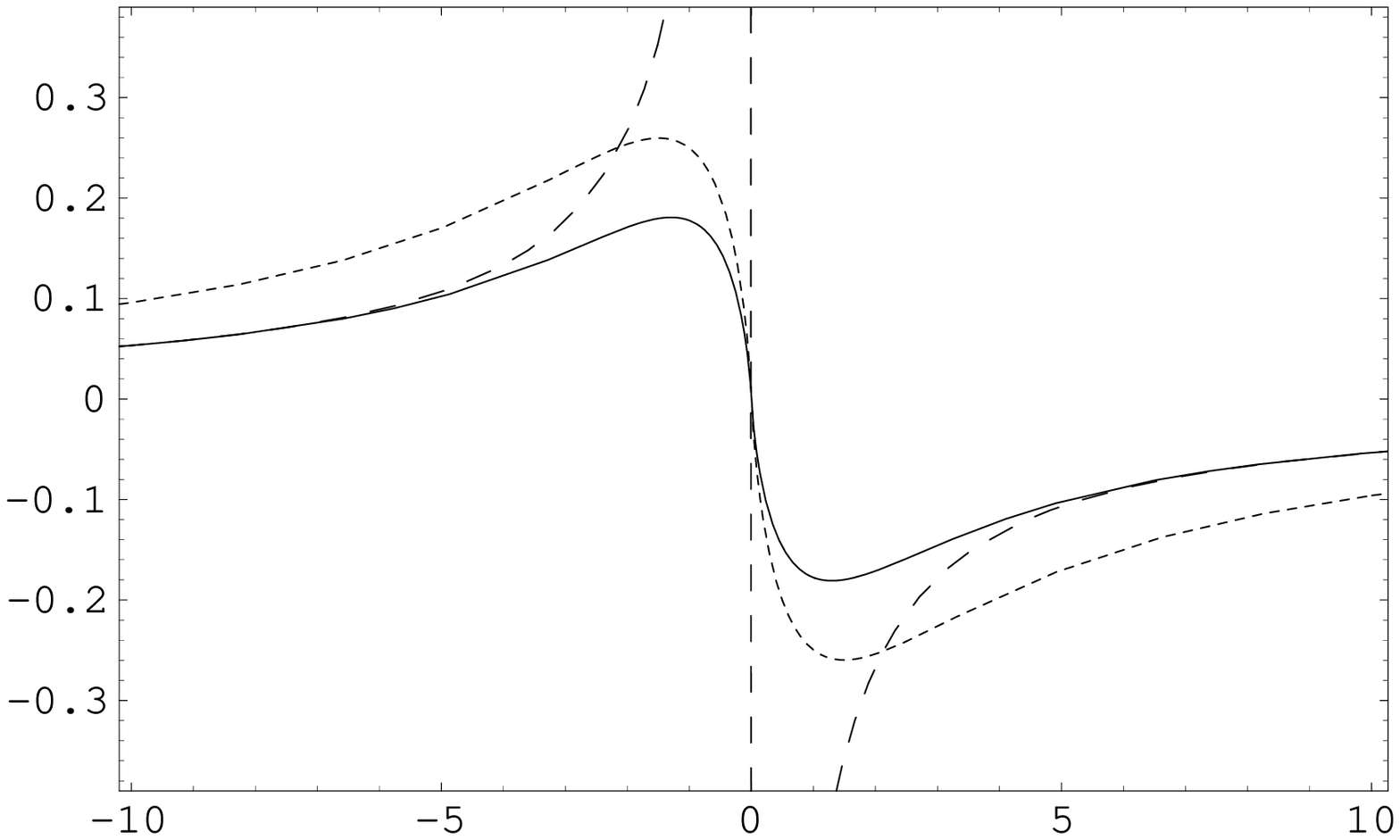,width=8cm}}
\put(4.0,0.0){$y/\ell_2$}
\put(-1.5,2.5){$\sigma_{yy}(0,y)$}
\end{picture}}
\vspace*{-1.0cm}
\centerline{
(d)\begin{picture}(8,6)
\put(0.0,0.2){\epsfig{file=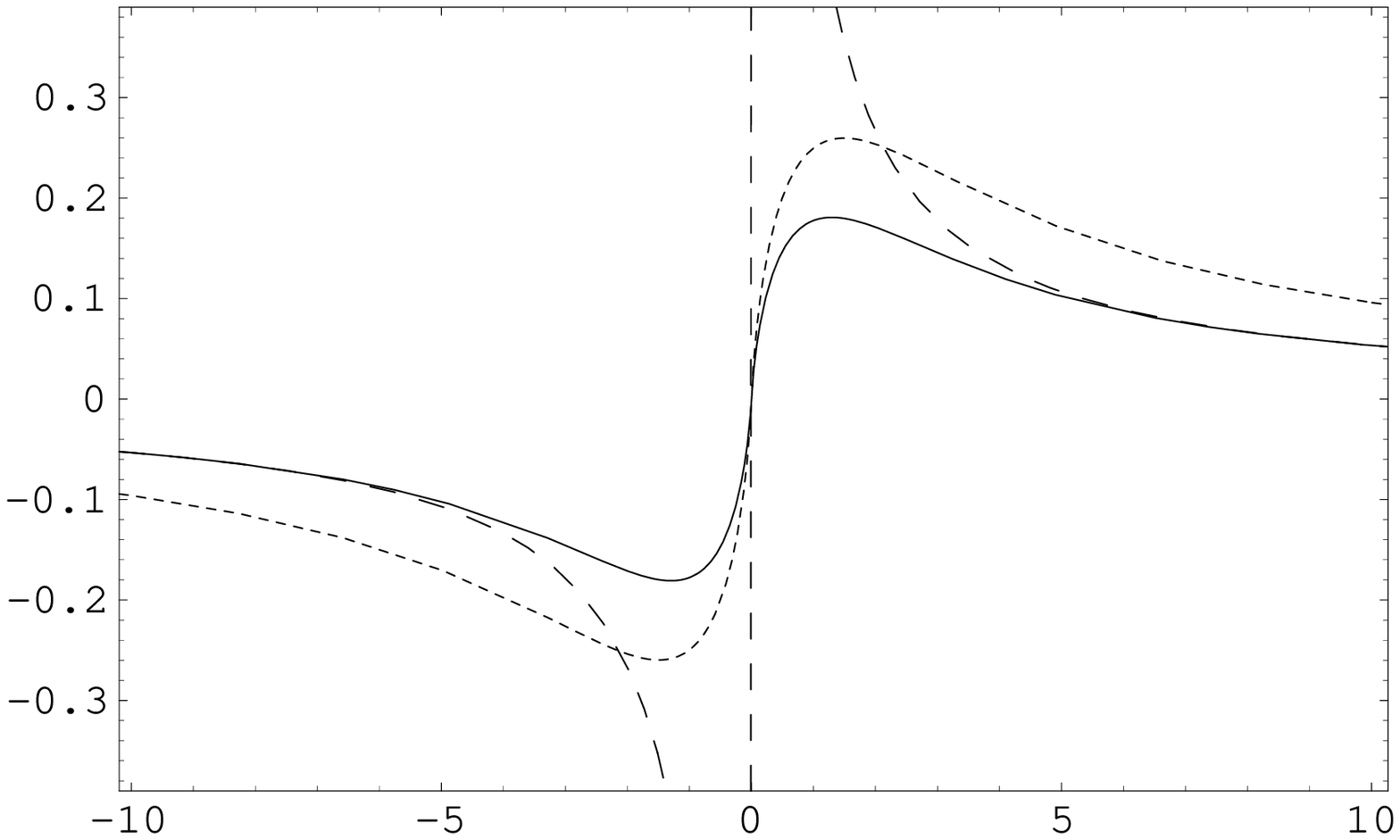,width=8cm}}
\put(4.0,0.0){$x/\ell_2$}
\put(-1.5,2.5){$\sigma_{yx}(x,0)$}
\end{picture}}
\caption{The force stresses components near the dislocation line: 
(a) $\sigma_{xx}(0,y)$, (b) $\sigma_{xy}(x,0)$, 
(c) $\sigma_{yy}(0,y)$, (d) $\sigma_{yx}(x,0)$ are given in units of 
$A$ with $\nu=0.3$ and $\gamma=\mu/2$. 
The dashed curves represent the stresses in asymmetric elasticity and 
the small dashed curves the symmetric force stresses ($\gamma=0$).}
\label{fig:stress-1D}
\end{figure}

If we substitute the stress functions~(\ref{f}) and (\ref{psi}) into the stress function ansatz~(\ref{SFA-edge}), the following components of the force stress tensor follow
\begin{align}
\label{Txx}
\sigma_{xx}&= -\frac{y}{r^4}\bigg\{A\Big[(y^2 + 3 x^2)
+\frac{4\,\ell^2_2}{r^2}(y^2 - 3x^2)
-2y^2\frac{r}{\ell_2}K_1\Big(\frac{r}{\ell_2}\Big)
- 2(y^2 - 3x^2)K_2\Big(\frac{r}{\ell_2}\Big)\Big]
\nonumber\\
&\quad
- B\Big[(x^2 - y^2) - \frac{4\,\ell^2_3}{r^2}(3x^2 -y^2)
+ 2x^2\frac{r}{\ell_3}K_1\Big(\frac{r}{\ell_3}\Big)
- 2(y^2 -3x^2)K_2\Big(\frac{r}{\ell_3}\Big)\Big]\bigg\}\\
\label{Tyy}
\sigma_{yy}&= -\frac{y}{r^4}\bigg\{ A\Big[(y^2 - x^2)
-\frac{4\,\ell^2_2}{r^2}(y^2 - 3x^2)
-2x^2\frac{r}{\ell_2}K_1\Big(\frac{r}{\ell_2}\Big)
+ 2(y^2 -3x^2)K_2\Big(\frac{r}{\ell_2}\Big)\Big]
\nonumber\\
&\quad + B\Big[(x^2 - y^2) - \frac{4\,\ell^2_3}{r^2}(3x^2 -y^2)
+ 2x^2\frac{r}{\ell_3}K_1\Big(\frac{r}{\ell_3}\Big)
+ 2(3x^2 -y^2)K_2\Big(\frac{r}{\ell_3}\Big)\Big]\bigg\}\\                             \label{Txy}
\sigma_{xy}&= \frac{x}{r^4}\bigg\{A\Big[(x^2 - y^2)
-\frac{4\,\ell^2_2}{r^2}(x^2 - 3y^2)
-2y^2\frac{r}{\ell_2}K_1\Big(\frac{r}{\ell_2}\Big)
+ 2(x^2 -3y^2)K_2\Big(\frac{r}{\ell_2}\Big)\Big]
\nonumber\\
&\quad
+ B\Big[(x^2 + 3y^2) + \frac{4\,\ell^2_3}{r^2}(x^2 -3y^2)
- 2x^2\frac{r}{\ell_3}K_1\Big(\frac{r}{\ell_3}\Big)
- 2(x^2 -3y^2)K_2\Big(\frac{r}{\ell_3}\Big)\Big]\bigg\}\\
\label{Tyx}
\sigma_{yx}&=\frac{x}{r^4}\bigg\{A\Big[(x^2 - y^2)
-\frac{4\,\ell^2_2}{r^2}(x^2 - 3y^2)
-2y^2\frac{r}{\ell_2}K_1\Big(\frac{r}{\ell_2}\Big)
+ 2(x^2 -3y^2)K_2\Big(\frac{r}{\ell_2}\Big)\Big]
\nonumber\\
&\quad
- B\Big[(x^2 - y^2) - \frac{4\,\ell^2_3}{r^2}(x^2 -3y^2)
- 2y^2\frac{r}{\ell_3}K_1\Big(\frac{r}{\ell_3}\Big)
+ 2(x^2 -3y^2)K_2\Big(\frac{r}{\ell_3}\Big)\Big]\bigg\}\\
\label{Tzz}
\sigma_{zz}&= - 2\nu\, A \,\frac{y}{r^2}
\Big[1 -\frac{r}{\ell_2}K_1\Big(\frac{r}{\ell_2}\Big)\Big]
\end{align}
with
\begin{align}
\label{}
A:=\frac{\mu b}{2\pi(1-\nu)},\qquad
B:=\frac{\mu \gamma b}{\pi(\mu + \gamma)}.
\end{align}
The trace of the stress tensor is
\begin{align}
\label{Tkk}
\sigma_{kk}&= - 2 (1 + \nu)\, A \,\frac{y}{r^2}
\Big[1 -\frac{r}{\ell_2}K_1\Big(\frac{r}{\ell_2}\Big)\Big].
\end{align}
The skew-symmetric part of the force stress tensor reads
\begin{align}
\label{Tskew}
\sigma_{[xy]}= \frac{\mu\gamma b}{\pi(\mu+\gamma)} \,
\frac{x}{r^2}\Big[1 -\frac{r}{\ell_3}K_1\Big(\frac{r}{\ell_3}\Big)\Big].
\end{align}
Now we want to discuss some details of the core modification of the force 
stress fields~(\ref{Txx})--(\ref{Tzz}).
The spatial distributions of the force stresses 
near the dislocation line are presented in figure~\ref{fig:stress-Co}.
The force stress fields have no artificial singularities at the core  
and the maximum stress occurs at a short distance away from 
the dislocation line (see figures~\ref{fig:stress-3D} and \ref{fig:stress-1D}). 
They are zero at $r=0$.
It can be seen that the stresses have the
following extreme values:
$|\sigma_{xx}(0,y)|\simeq 0.546 A/\ell_2+0.260 B/\ell_3$ at 
$|y|\simeq (0.996 \ell_2+1.494\ell_3)/2$,
$|\sigma_{yy}(0,y)|\simeq 0.260 A/\ell_2-0.260 B/\ell_3$ at 
$|y|\simeq (1.494 \ell_2+1.494\ell_3)/2$,
$|\sigma_{xy}(x,0)|\simeq 0.260 A/\ell_2+0.546 B/\ell_3$ at 
$|x|\simeq (1.494 \ell_2+0.996 \ell_3)/2$,
$|\sigma_{yx}(x,0)|\simeq 0.260 A/\ell_2-0.260 B/\ell_3$ at 
$|x|\simeq (1.494 \ell_2+1.494\ell_3)/2$,
and
$|\sigma_{zz}(0,y)|\simeq 0.399 A$ at 
$|y|\simeq 1.114 \ell_2$.
Thus, the characteristic internal lengths $\ell_2$ and $\ell_3$ determine the
position and the magnitude of the stress maxima.
For $\gamma>0$ the stresses $\sigma_{xx}$ and $\sigma_{xy}$ are bigger
than in the case $\gamma=0$  and $\sigma_{yy}$ and $\sigma_{yx}$ are smaller
than in the case $\gamma=0$ (see figure~\ref{fig:stress-1D}).

Using Eq.~(\ref{CR-B}), we find for the elastic distortion
\begin{align}
\label{Bxx}
\beta_{xx}&=-\frac{y}{r^4}\bigg\{\frac{A}{2\mu}\Big[(1-2\nu)r^2 + 2x^2
+\frac{4\,\ell^2_2}{r^2}(y^2 - 3x^2)\\
&\hspace{3cm}
- 2(y^2 - \nu\,r^2) \frac{r}{\ell_2}K_1\Big(\frac{r}{\ell_2}\Big)
- 2(y^2 -3x^2)K_2\Big(\frac{r}{\ell_2}\Big)\Big]
\nonumber\\
& \hspace{1cm}
- \frac{B}{2\mu}\Big[(x^2 - y^2) - \frac{4\,\ell^2_3}{r^2}(3x^2 -y^2)
+ 2x^2\frac{r}{\ell_3}K_1\Big(\frac{r}{\ell_3}\Big)
- 2(y^2 -3x^2)K_2\Big(\frac{r}{\ell_3}\Big)\Big]\bigg\}\nonumber\\
\label{Byy}
\beta_{yy}&=-\frac{y}{r^4}\bigg\{\frac{A}{2\mu}\Big[(1-2\nu)r^2 - 2x^2 -
\frac{4\,\ell^2_2}{r^2}(y^2 - 3x^2)\\
& \hspace{3cm}
-2(x^2 - \nu\,r^2)\frac{r}{\ell_2}K_1\Big(\frac{r}{\ell_2}\Big)
+ 2(y^2 -3x^2)K_2\Big(\frac{r}{\ell_2}\Big)\Big]
\nonumber\\
&\hspace{1cm}
+ \frac{B}{2\mu}\Big[(x^2 - y^2) - \frac{4\,\ell^2_3}{r^2}(3x^2 -y^2)
+ 2x^2\frac{r}{\ell_3}K_1\Big(\frac{r}{\ell_3}\Big)
+ 2(3x^2 -y^2)K_2\Big(\frac{r}{\ell_3}\Big)\Big]\bigg\}
\nonumber\\
\label{Bxy}
\beta_{xy}&=\frac{x}{r^4}\bigg\{\frac{A}{2\mu}\Big[(x^2 - y^2)
-\frac{4\,\ell^2_2}{r^2}(x^2 - 3y^2)
-2y^2\frac{r}{\ell_2}K_1\Big(\frac{r}{\ell_2}\Big) + 2(x^2 -
3y^2)K_2\Big(\frac{r}{\ell_2}\Big)\Big]
\nonumber\\
&\hspace{1cm}
+ \frac{B}{2\mu}\Big[2y^2 + \frac{4\,\ell^2_3}{r^2}(x^2 -3y^2)
- (x^2 - y^2)\frac{r}{\ell_3}K_1\Big(\frac{r}{\ell_3}\Big)
- 2(x^2 -3y^2)K_2\Big(\frac{r}{\ell_3}\Big)\Big]
\nonumber\\
&\hspace{1cm}
+ \frac{B}{2\gamma}\,r^2\Big[1 -\frac{r}{\ell_3}K_1\Big(\frac{r}{\ell_3}\Big)\Big]\bigg\}\\
\label{Byx}
\beta_{yx}&=\frac{x}{r^4}\bigg\{\frac{A}{2\mu}\Big[(x^2 - y^2)
-\frac{4\,\ell^2_2}{r^2}(x^2 - 3y^2)
-2y^2\frac{r}{\ell_2}K_1\Big(\frac{r}{\ell_2}\Big)
+ 2(x^2 -3y^2)K_2\Big(\frac{r}{\ell_2}\Big)\Big]
\nonumber\\
&\hspace{1cm}
+ \frac{B}{2\mu}\Big[2y^2 + \frac{4\,\ell^2_3}{r^2}(x^2 -3y^2)
- (x^2 - y^2)\frac{r}{\ell_3}K_1\Big(\frac{r}{\ell_3}\Big)
- 2(x^2 -3y^2)K_2\Big(\frac{r}{\ell_3}\Big)\Big]
\nonumber\\
&\hspace{1cm}
-\frac{B}{2\gamma}\,r^2\Big[1 - \frac{r}{\ell_3}K_1\Big(\frac{r}{\ell_3}\Big)\Big] \bigg\}.
\end{align}
The dilatation reads
\begin{align}
\label{Bkk}
\beta_{kk}=-  \frac{(1 -2 \nu) A}{\mu} \,\frac{y}{r^2}\Big[1 -
\frac{r}{\ell_2}K_1\Big(\frac{r}{\ell_2}\Big)\Big].
\end{align}
In addition the elastic rotation is
\begin{align}
\label{Bskew}
\beta_{[xy]}= \frac{\mu b}{2\pi(\mu+\gamma)} \,
\frac{x}{r^2}\Big[1 -\frac{r}{\ell_3}K_1\Big(\frac{r}{\ell_3}\Big)\Big].
\end{align}
In Eqs.~(\ref{Bkk}) and (\ref{Bskew}) it can be seen
that $\ell_2$ and $\ell_3$ are the characteristic lengths for
the elastic dilatation and elastic rotation, respectively.

\begin{figure}[t]\unitlength1cm
\centerline{
\epsfig{figure=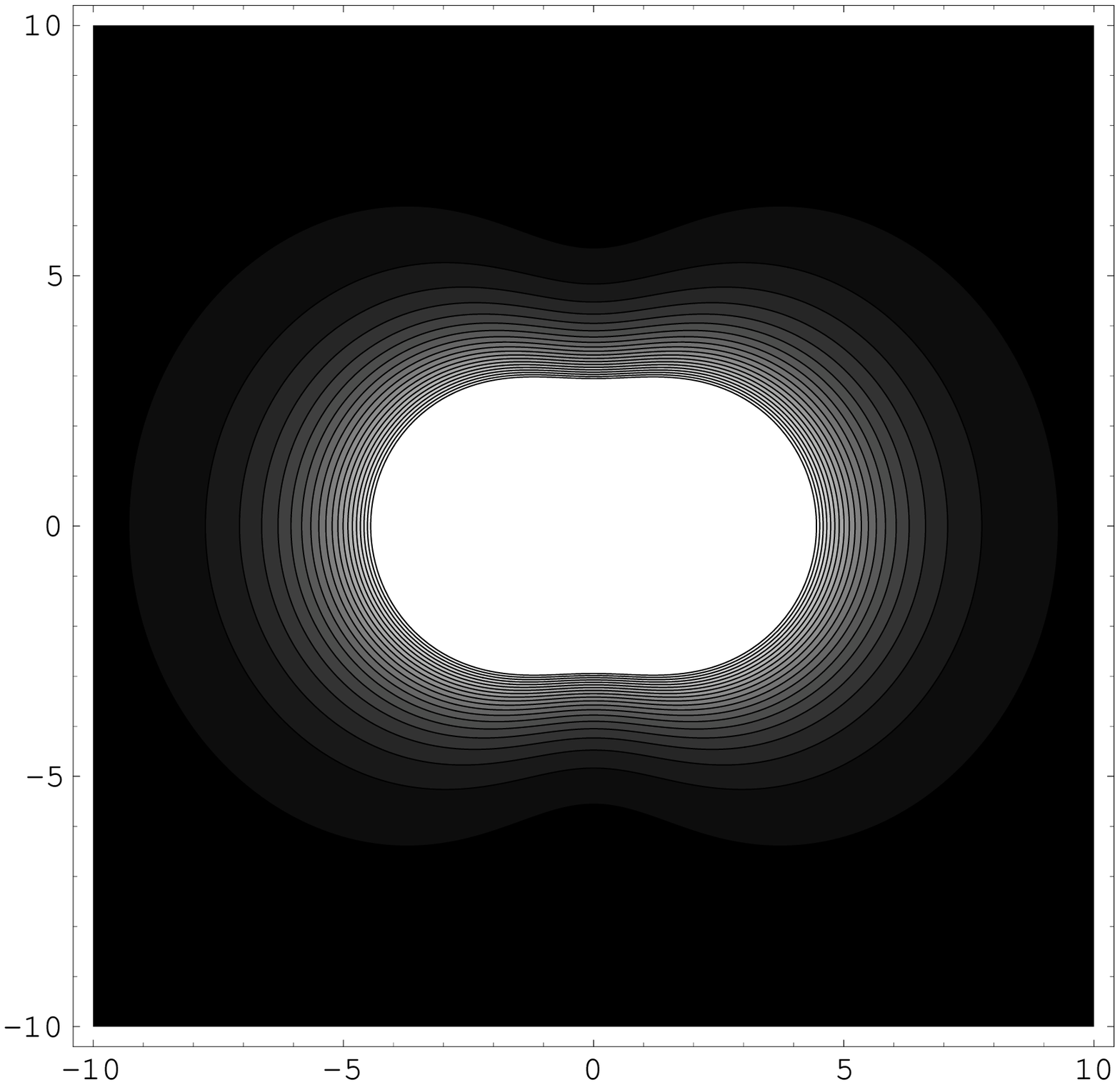,width=6.0cm}
\put(-6.7,3.0){$y/\ell_2$}
\put(-3.0,-0.3){$x/\ell_2$}
\put(-6.2,-0.3){$\text{(a)}$}
\hspace*{0.2cm}
\put(0,-0.3){$\text{(b)}$}
\epsfig{figure=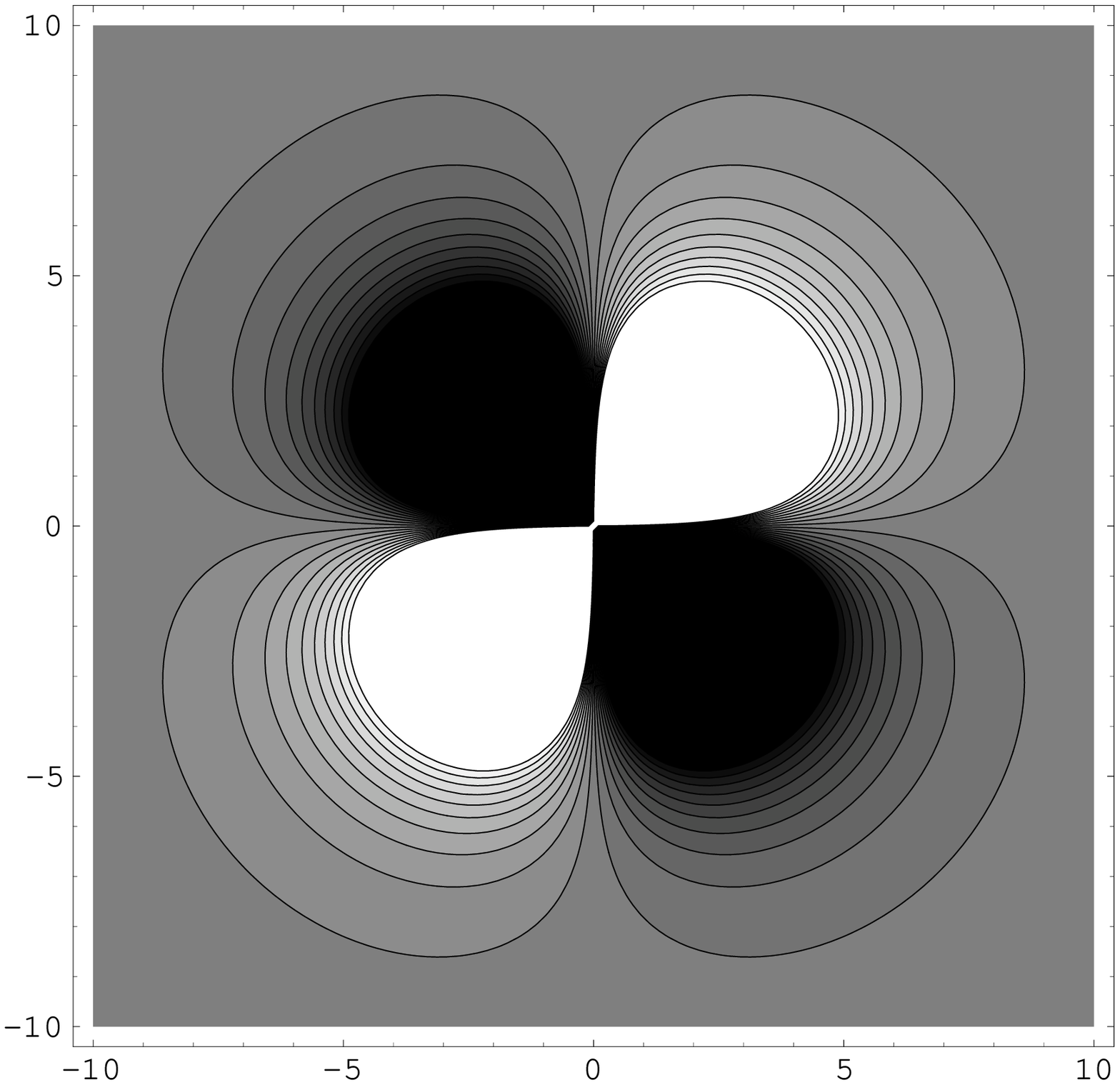,width=6.0cm}
\put(-3.0,-0.3){$ x/\ell_2$}
}
\caption{Torsion contours of an edge dislocation near the dislocation line:
(a) $\alpha_{xz}$, (b) $\alpha_{yz}$ with $\nu=0.3$ and $\gamma=\mu/2$ (in units 
of $b/[4\pi]$).}
\label{fig:DD}
\end{figure}

Using the elastic distortion~(\ref{Bxx})--(\ref{Byx}) in terms of the stress functions $f$ and $\Psi$, we obtain 
for the torsion~(\ref{Tor}) or the dislocation
density of a straight edge dislocation according
\begin{align}
\alpha_{xz}&=T_{xxy}=-\frac{1-\nu}{2\mu}\, \pd_y \Delta f
+\frac{\mu+\gamma}{4\mu\gamma}\, \pd_x \Delta \Psi\\
\alpha_{yz}&=T_{yxy}=\frac{1-\nu}{2\mu}\, \pd_x \Delta f
+\frac{\mu+\gamma}{4\mu\gamma}\, \pd_y \Delta \Psi.
\end{align}
Simple differentiation gives the non-vanishing expressions
\begin{align}
\label{DD-edge-x}
\alpha_{xz}&=\frac{b}{4\pi}\bigg\{
\frac{1}{\ell^2_2}\,K_0\Big(\frac{r}{\ell_2}\Big) +
\frac{1}{\ell^2_3}\,K_0\Big(\frac{r}{\ell_3}\Big)
- \frac{x^2-y^2}{r^2}
\Big[\frac{1}{\ell^2_2}\,K_2\Big(\frac{r}{\ell_2}\Big) -
\frac{1}{\ell^2_3}\,K_2\Big(\frac{r}{\ell_3}\Big)\Big]\bigg\} \\
\label{DD-edge-y}
\alpha_{yz}&=-\frac{b}{2\pi}
\frac{x y}{r^2}
\Big[\frac{1}{\ell^2_2}\,K_2\Big(\frac{r}{\ell_2}\Big) -
\frac{1}{\ell^2_3}\,K_2\Big(\frac{r}{\ell_3}\Big)\Big].
\end{align}
So far, it is  surprising that the component~(\ref{DD-edge-y}),
which is usually the dislocation density of an edge dislocation with Burgers
vector $b_y$, is non-zero. 
The components~(\ref{DD-edge-x}) and (\ref{DD-edge-y}) are necessary to fulfill
the pseudomomentum equilibrium condition~(\ref{ME}).
Also we would like to note that these non-vanishing components of the torsion tensor
do not have cylindrical symmetry due to the $K_2$-terms (see figure~\ref{fig:DD}).
Since an edge dislocation is lacking cylindrical symmetry around the
dislocation line 
two length scales, $\ell_2$ and $\ell_3$, 
are needed for a proper model.
Fig.~\ref{fig:DD}a shows this asymmetry in the dislocation core region of an edge dislocation. 

\begin{figure}[t]\unitlength1cm
\vspace*{-0.5cm}
\centerline{
\begin{picture}(8,6)
\put(0.0,0.2){\epsfig{file=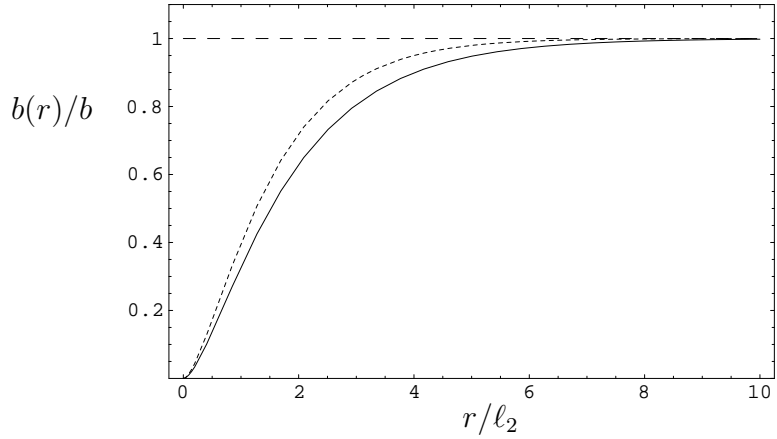,width=9cm}}
\put(4.5,-0.1){$r/\ell_2$}
\put(-1.5,4.0){$b(r)/b$}
\end{picture}
}
\caption{The modified Burgers vector of an edge dislocation $b(r)/b$
for $\ell_2=\ell_3$ (small dashed) and for $\nu=0.3$ and $\gamma=\mu/2$ (solid).}
\label{fig:Burger-edge}
\end{figure}

With Eqs.~(\ref{moment1}) and (\ref{moment3})
the localized pseudomoment  stresses of bending type are given by
\begin{align}
H_{xz}&=H_{xxy}=\frac{1}{1-\nu}\, (c_1-c_2) \alpha_{xz}\\
      &=
A\Big[K_0\Big(\frac{r}{\ell_2}\Big)
-\frac{x^2-y^2}{r^2}\, K_2\Big(\frac{r}{\ell_2}\Big)\Big]
+ B
\Big[K_0\Big(\frac{r}{\ell_3}\Big)
+\frac{x^2-y^2}{r^2}\, K_2\Big(\frac{r}{\ell_3}\Big)\Big] \Big\}\nonumber\\
H_{zx}&=H_{zxy}=-\frac{\nu}{1-\nu}\, (c_1-c_2) \alpha_{xz}\\
      &=-\nu A
\Big[K_0\Big(\frac{r}{\ell_2}\Big)
-\frac{x^2-y^2}{r^2}\, K_2\Big(\frac{r}{\ell_2}\Big)\Big]
-\nu B
\Big[K_0\Big(\frac{r}{\ell_3}\Big)
+\frac{x^2-y^2}{r^2}\, K_2\Big(\frac{r}{\ell_3}\Big)\Big] \Big\}\nonumber\\
H_{yz}&=H_{yxy}=\frac{1}{1-\nu}\, (c_1-c_2) \alpha_{yz}
      =
-2\, \frac{xy}{r^2}\Big[ A\, K_2\Big(\frac{r}{\ell_2}\Big)
- B K_2\Big(\frac{r}{\ell_3}\Big)\Big]\\
H_{zy}&=H_{zzx}=-\frac{\nu}{1-\nu}\, (c_1-c_2) \alpha_{yz}
      =2 \nu\, \frac{xy}{r^2} \Big[ A\, K_2\Big(\frac{r}{\ell_2}\Big)
- B K_2\Big(\frac{r}{\ell_3}\Big)\Big] .
\end{align}
Therefore, plane strain bending is given in terms of the two length scales
$\ell_2$ and $\ell_3$.
If we use the components~(\ref{DD-edge-x}) and (\ref{DD-edge-y})
 of the dislocation density, the Burgers vector is calculated as
\begin{align}
\label{Burger-edge-x}
b(r)&=\oint (\beta_{xx}\,\d x + \beta_{xy}\, \d y) = \int^{2\pi}_0\int^r_0 \alpha_{xz}(r',\phi')\,r'\,\d r'\,\d\phi'
\nonumber\\
&= b\Big\{1-\frac{1}{2}\,\Big[\frac{r}{\ell_2}
\,K_1\Big(\frac{r}{\ell_2}\Big) + \frac{r}{\ell_3}
\,K_1\Big(\frac{r}{\ell_3}\Big)\Big]\Big\}\\
\label{Burger-edge-y}
0&=\oint (\beta_{yx}\,\d x + \beta_{yy}\, \d y) 
= \int^{2\pi}_0\int^r_0 \alpha_{yz}(r',\phi')\,r'\,\d r'\,\d\phi' .
\end{align}
Thus, it can be seen that the 
torsion~(\ref{DD-edge-y}) does not contribute
to the Burgers vector. Only the $K_0$-terms in (\ref{DD-edge-x}) give a contribution
to the Burgers vector~(\ref{Burger-edge-x}).
The plot is given in figure~\ref{fig:Burger-edge}.

\subsection{The case $\ell_2=\ell_3$}
Now we turn to a special case of the gauge theoretical edge
dislocation we have found. We obtained two internal lengths $\ell_2$
and $\ell_3$. 
If we require that the dislocation density~(\ref{DD-edge-y}) has to be zero,
the internal lengths must fulfill $\ell_2=\ell_3$.
Therefore, we may reduce them to only one internal length. If
the internal lengths $\ell_2$ and $\ell_3$ are equal, we find
\begin{align}
\ell_2=\ell_3:\qquad \Rightarrow\qquad
 \gamma=\frac{\lambda}{2\nu}=\frac{\mu}{1-2\nu},\qquad
 A=B.
\end{align}
Thus, in this case the parameter $\gamma$ is expressed in terms of
$\mu$ and $\nu$. 
Eventually, the non-vanishing components of the stress tensor are
\begin{align}
\label{Txx-V} \sigma_{xx}&= -\frac{\mu
b}{\pi(1-\nu)}\frac{y}{r^2}\Big[1-\frac{r}{\ell_2}K_1\Big(\frac{r}{\ell_2}\Big)\Big]
\\
\label{Txy-V}
\sigma_{xy}&= \frac{\mu b}{\pi(1-\nu)}\frac{x}{r^2}\Big[1-\frac{r}{\ell_2}K_1\Big(\frac{r}{\ell_2}\Big)\Big]\\
\label{Tzz-V} \sigma_{zz}&= -\frac{\mu\nu
b}{\pi(1-\nu)}\frac{y}{r^2}\Big[1-\frac{r}{\ell_2}K_1\Big(\frac{r}{\ell_2}\Big)\Big].
\end{align}
Thus, $\sigma_{yx}=0$ and $\sigma_{yy}=0$. 
The non-vanishing components of the elastic distortion are
\begin{align}
\label{Bxx-V}
\beta_{xx}&= -\frac{b}{2\pi }\frac{y}{r^2}\Big[1-\frac{r}{\ell_2}K_1\Big(\frac{r}{\ell_2}\Big)\Big]\\
\label{Bxy-V}
\beta_{xy}&= \frac{b}{2\pi}\frac{x}{r^2}\Big[1-\frac{r}{\ell_2}K_1\Big(\frac{r}{\ell_2}\Big)\Big]\\
\label{Byx-VV} \beta_{yx}&= \frac{\nu
b}{2\pi(1-\nu)}\frac{x}{r^2}\Big[1-\frac{r}{\ell_2}K_1\Big(\frac{r}{\ell_2}\Big)\Big]\\
\label{Bzz-VV} \beta_{yy}&= \frac{\nu
b}{2\pi(1-\nu)}\frac{y}{r^2}\Big[1-\frac{r}{\ell_2}K_1\Big(\frac{r}{\ell_2}\Big)\Big].
\end{align}
Because the components $\beta_{yx}$ and $\beta_{yy}$ are
compatible, they can be expressed in terms of a displacement field
$u_y$ according to
\begin{align}
\beta_{yx}=u_{y,x},\qquad \beta_{yy}=u_{y,y},\qquad
u_{y}=\frac{\nu b}{2\pi(1-\nu)}\,
\Big[\ln r+K_0\Big(\frac{r}{\ell_2}\Big)\Big].
\end{align}
Let us compare this type of edge dislocation with the edge
dislocation originally introduced by~\citet{Volterra} (see also~\citep{deWit73}). 
The displacement
field of a Volterra edge dislocation reads $u_x=b/(2\pi) \arctan y/x$
and $u_y=b/(2\pi) \ln r$. But the corresponding force stress of
this displacement field calculated in the theory of (symmetric)
elasticity does not fulfill the force equilibrium condition
because it produces line forces at the dislocation line and the
dilatation and the hydrostatic pressure are zero. These drawbacks
are not acceptable from the physical point of view. In the limit
$\ell_2 \rightarrow 0$, our displacement field $u_y$ has a
pre-factor $\nu/(1-\nu)$ instead of 1. But the modified pre-factor
is necessary to satisfy the force equilibrium condition with
asymmetric force stresses and to give the correct dilatation.
Thus, we have found the gauge theoretical version of an edge
dislocation of Volterra type which is able to remove the original
drawbacks of the edge dislocation of Volterra type.
The limiting value $\nu=1/2$ defines incompressibility and 
we obtain the pre-factor 1. 
In this limit $\lambda$ and $\gamma$ tend to infinity. 
Thus, in the limit of incompressibility 
the Volterra dislocation is also valid.

The non-vanishing component of the dislocation density tensor is 
given by
\begin{align}
\label{DD-edge-V} 
\alpha_{xz}=
\frac{b}{2\pi}\frac{1}{\ell^2_2}\,K_0\Big(\frac{r}{\ell_2}\Big).
\end{align}
Thus, this dislocation density has cylindrical symmetry and has the same form
as a screw dislocation~(\ref{Tor-screw}).
The localized pseudomoment stress reads
\begin{align}
\label{H-V} 
H_{xz}= \frac{\mu b}{\pi(1-\nu)}\,K_0\Big(\frac{r}{\ell_2}\Big),\qquad
H_{zx}=-\frac{\nu\mu b}{\pi(1-\nu)}\,K_0\Big(\frac{r}{\ell_2}\Big).
\end{align}
It is the pseudomoment stress of an edge dislocation with cylindrical symmetry (see also~\citep{Lazar03}).
The corresponding Burgers vector is
\begin{align}
b(r) =
b\Big[1-\frac{r}{\ell_2}\,K_1\Big(\frac{r}{\ell_2}\Big)\Big].
\end{align}

\subsection{The case $\gamma=0$: symmetric force stresses}
\label{g0}
Now we want to revert the case of symmetric force stresses.
If we set $\gamma=0$ (symmetric force stress), we obtain 
from Eqs.~(\ref{Rel-screw}), (\ref{L1}), (\ref{Rel-edge}), (\ref{L2}) 
and (\ref{L3}):
\begin{align}
\label{Rel-sy}
a_3=-\frac{a_1}{2},\qquad a_2=\frac{1+\nu}{1-\nu}\, a_1, 
\qquad \ell_2^2=\ell_1^2= \frac{a_1}{2\mu},\qquad\ell_3^2\rightarrow\infty,
\qquad B=0.
\end{align}
It is obvious that $a_3$ in (\ref{Rel-sy}) violates the 
condition of positive dislocation core energy (\ref{EC-a}).
Thus, the condition of nonnegative dislocation core energy 
requires asymmetric force stresses.
The condition (\ref{Rel-sy}) has only one independent internal length and 
it is the choice introduced by~\citet{Lazar03}.
With $B=0$ we recover from Eqs.~(\ref{Txx})--(\ref{Tzz}) the force stress earlier calculated by~\cite{Lazar03} and~\cite{GA99,LM05}
in the framework of a gauge theory with symmetric force stresses and strain gradient
elasticity, respectively.
Also the distortions~(\ref{Bxx})--(\ref{Byx}) become symmetric with $B=0$
and the elastic rotation~(\ref{Bskew}) becomes zero. 


\subsection{The case $\gamma\rightarrow\infty$}
If we set $\gamma\rightarrow \infty$ , then we obtain 
from Eqs.~(\ref{Rel-screw}), (\ref{L1}), (\ref{Rel-edge}), (\ref{L2}) 
and (\ref{L3}):
\begin{align}
\label{Rel-inf}
a_3\rightarrow\infty,\qquad a_2=\frac{1+\nu}{1-\nu}\, a_1, 
\qquad \ell_2^2=\ell_1^2= \frac{a_1}{2\mu},\qquad\ell_3^2=\frac{a_1}{4\mu(1-\nu)},
\qquad B=\frac{\mu b}{\pi}.
\end{align}
This case does not violate the positive definiteness of the dislocation core
energy (\ref{EC-a}).
So we obtain the relation between $\ell_2$ and $\ell_3$
\begin{align}
\ell_3^2=\frac{1}{2(1-\nu)}\,\ell_2^2,\qquad 
\frac{\ell_2}{2}\le \ell_3\le \ell_2.
\end{align}
The skewsymmetric stress~(\ref{Tskew}) converts to
\begin{align}
\label{Tskew-3}
\sigma_{[xy]}= \frac{\mu b}{\pi} \,
\frac{x}{r^2}\Big[1 -\frac{r}{\ell_3}K_1\Big(\frac{r}{\ell_3}\Big)\Big]
\end{align}
and the  skewsymmetric distortion~(\ref{Bskew}) vanishes
$\beta_{[xy]}=0$.
All other quantities can be easily calculated with (\ref{Rel-inf}).
Because the elastic rotation is zero 
in the present case 
the corresponding energy converts to
\begin{align}
\lim_{\gamma\rightarrow\infty}W(\Bbeta,{\text{curl}}\,\Bbeta)
=W(\Be,{\text{curl}}\,\Be)
=W(\Be,{\text{curl}}\,\Be^{\text{P}}),
\qquad \Be={\text{sym}}\,\Bbeta
\end{align}
where $\Be$ and $\Be^{\text{P}}$ denote the elastic and plastic strains. 
Such a theory looks like a theory of strain gradient plasticity (see, e.g.,~\citet{Gurtin05}.)
In such a theory the trace of the dislocation density tensor is zero
\begin{align}
\alpha_{jj}=-\epsilon_{jkl} e_{jk,l}=
\epsilon_{jkl} e_{jk,l}^{\text{P}}=0.
\end{align}
Thus, it is a theory with two constraints
\begin{align}
\gamma\rightarrow\infty:&\qquad \beta_{[ij]}=0\nonumber\\
a_3\rightarrow\infty:&\qquad \alpha_{jj}=0 .
\end{align}

\section{Conclusion}
\setcounter{equation}{0}
\label{concl}
In this paper, we have investigated the (static) 
translational gauge theory of dislocations. 
We have used the most general linear and isotropic constitutive relations 
between the 
force stress and elastic distortion tensors and the pseudomoment stress and 
dislocation density (torsion) tensors. 
Thus, the linear theory possesses six material coefficients
and the force and pseudomoment stress tensors are asymmetric.
We have found four characteristic length scales for the gauge theory of dislocation
given in terms of the six material coefficients to account for size effects of small-scale problems.
We have derived the conditions of positive energy for the six material coefficients.
We observed that the so-called Einstein choice of the three coefficients of the 
pseudomoment stress tensor violates the positive definiteness of the dislocation core energy.
This fact demonstrates that the three-dimensional Einsteinian gauge approach is not a
realistic model for dislocations. 
We have observed that in the gauge theory of dislocations
moment stress is the response to contortion and pseudomoment stress is the response
to pseudomoment stress.
Using Fourier transform, we have calculated the three-dimensional 
Green tensor of our so-called 
master equation of the force stress tensor.
In the case of generalized plane strain the four characteristic lengths 
reduce to only two lengths and the number of independent material coefficients 
simplifies to four.
Later, we have solved the anti-plane strain problem of a screw dislocation and
the plane strain problem of an edge dislocation in the framework of gauge theory.
In this turn, we have found new solutions for all physical state quantities of
a screw dislocation and an edge dislocation.

\section*{Acknowledgment}
The authors have been supported by an Emmy-Noether grant of the 
Deutsche Forschungsgemeinschaft (Grant No. La1974/1-2). 
One of us (M.L.) is very grateful to Friedrich W. Hehl for very useful and stimulating
discussions about the pseudomoment stress tensor and the translational gauge theory
in general.

\begin{appendix}
\setcounter{equation}{0}
\renewcommand{\theequation}{\thesection.\arabic{equation}}
\section{Screw dislocation in asymmetric elasticity}
\label{appendixA}

In asymmetric elasticity the force stress tensor is asymmetric and fulfills the
force equilibrium condition
\begin{align}
\label{FE-0}
\sigma^0_{ij,j}=0,\qquad \sigma^0_{ij}\neq \sigma_{ji}^0.
\end{align}
The elastic distortion of a Volterra screw dislocation is given by~\cite{deWit73}
\begin{align}
\label{B-0}
\beta^0_{zx}= -\frac{b}{2\pi}\frac{y}{r^2},\qquad\beta^0_{zy}= \frac{b}{2\pi}\frac{x}{r^2},\qquad r^2=x^2+y^2.
\end{align}
Thus, the present problem is of anti-plane strain type.
The non-vanishing component of the dislocation density of a screw dislocation
reads
\begin{align}
\label{DD-screw0}
\alpha^0_{zz} = \beta^0_{zy,x} - \beta^0_{zx,y}.
\end{align}
It is the incompatibility condition.
In terms of force stresses it reads
\begin{align}
\label{}
\alpha^0_{zz} = \frac{1}{\mu + \gamma}(\sigma^0_{zy,x} - \sigma^0_{zx,y}).
\end{align}
For a straight screw dislocation it has the form
\begin{align}
\label{}
\alpha^0_{zz} = b\,\delta{(x)}\delta{(y)}.
\end{align}

In terms of the Prandtl stress function $F^0$, the incompatibility condition
is given by
\begin{align}
\label{}
\Delta F^0 = (\mu + \gamma)b\,\delta{(x)}\delta{(y)}.
\end{align}
The Prandtl stress function is nothing but the Green function
of the two-dimensional Laplace equation. It reads~\citep{Kroener81}
\begin{align}
\label{F-0}
F^0 = \frac{(\mu + \gamma)b}{2\pi}\,\ln r.
\end{align}

\setcounter{equation}{0}
\renewcommand{\theequation}{\thesection.\arabic{equation}}
\section{Edge dislocation in asymmetric elasticity}
\label{appendixB}

In the case of straight edge dislocations the equations of incompatibility
take the form
\begin{align}
\label{}
\alpha^0_{xz}&=\beta^0_{xy,x} - \beta^0_{xx,y}\\
\label{}
\alpha^0_{yz}&=\beta^0_{yy,x} - \beta^0_{yx,y}.
\end{align}
We introduce the following combinations~\cite{Nowacki86}
\begin{align}
\label{A1}
A_1:=&\,\alpha^0_{yz,x} - \alpha^0_{xz,y} = \beta^0_{yy,xx} + \beta^0_{xx,yy} -
\beta^0_{xy,xy} -\beta^0_{yx,xy}\\
\label{A2}
A_2:=& - \alpha^0_{xz,x} - \alpha^0_{yz,y} = \beta^0_{yx,yy} - \beta^0_{xy,xx} + \beta^0_{xx,xy} - \beta^0_{yy,xy} .
\end{align}
Expressing the elastic distortions in terms of force stresses and using $\sigma^0_{ij,j}=0$ and 
$\sigma^0_{zz}=\nu(\sigma^0_{xx}+\sigma^0_{yy})$, we obtain
\begin{align}
\label{A1-S}
A_1=&\frac{1-\nu}{2\mu}\,\Delta(\sigma^0_{xx} + \sigma^0_{yy})\\
\label{A2-S}
A_2=& \frac{1}{2\mu}(\sigma^0_{xx,xy} - \sigma^0_{yy,xy}) +
\frac{\gamma - \mu}{4\mu\gamma}(\sigma^0_{xy,yy} - \sigma^0_{yx,xx}) +
\frac{\gamma + \mu}{4\mu\gamma}(\sigma^0_{yx,yy} - \sigma^0_{xy,xx}).
\end{align}

Because we deal with asymmetric force stresses we use the stress function ansatz
given by Mindlin for couple-stress theory~\cite{Mindlin63}
\begin{align}
\label{AM0}
\sigma^0_{ij}=
\left(\begin{array}{ccc}
\pd^2_{yy}f^0 - \pd^2_{xy}\Psi^0  & -\pd^2_{xy}f^0 + \pd^2_{xx}\Psi^0  & 0 \\\\
-\pd^2_{xy}f^0 - \pd^2_{yy}\Psi^0 & \pd^2_{xx}f^0 + \pd^2_{xy}\Psi^0  & 0 \\\\
 0  &  0  & \nu \Delta f^0
\end{array} \right)
\end{align}
where $f^0$ and $\Psi^0$ are stress functions of second order. 
The stress function ansatz~(\ref{AM0}) is the generalization of the 
stress function ansatz with the Airy stress function $f^0$ for symmetric
stresses. If $\Psi^0$ is zero, (\ref{AM0}) reduces to the usual expression for the stresses
in terms of the Airy stress function $f^0$.
Equations~(\ref{A1-S}) and (\ref{A2-S}) are reduced to the
following inhomogeneous bi-harmonic equations
\begin{align}
\label{}
\Delta\Delta\,f^0&= \frac{2\mu}{1-\nu}A_1\\
\label{}
\Delta\Delta\,\Psi^0&= - \frac{4\mu\gamma}{\mu + \gamma} A_2.
\end{align}
Because we want to consider a straight edge dislocation with the Burgers vector $b=b_x$,
the dislocation density tensor has the form
\begin{align}
\label{}
\alpha^0_{yz}=0,\qquad \alpha^0_{xz}=b\,\delta(x)\delta(y).
\end{align}
So we obtain
\begin{align}
\label{BH-f}
\Delta\Delta\,f^0&= - \frac{2\mu b}{1-\nu}\,\pd_y[\delta(x)\delta(y)]\\
\label{BH-psi}
\Delta\Delta\,\Psi^0&= \frac{4\mu\gamma b}{\mu + \gamma}\,\pd_x[\delta(x)\delta(y)].
\end{align}
Since the two-dimensional Green function of the bi-harmonic equation
is
\begin{align}
\label{}
\Delta\Delta\,G&= \delta(x)\delta(y),\qquad G= \frac{1}{8\pi}\,r^2\ln r
\end{align}
the solutions of (\ref{BH-f}) and (\ref{BH-psi}) are the following
Airy stress functions~\citep{Kroener81}
\begin{align}
\label{f0}
f^0&= - \frac{\mu b}{4\pi(1-\nu)}\, \pd_y (r^2 \ln r)\\
\label{psi0}
\Psi^0&= \frac{\mu\gamma\,b}{2\pi(\mu + \gamma)}\,\pd_x(r^2 \ln r).
\end{align}
(\ref{f0}) is the well-known Airy stress function for an edge
dislocation with Burgers vector $b_x$ and (\ref{psi0}) looks like
an Airy stress function for an edge dislocation with Burgers
vector $b_y$ with a different pre-factor.

\end{appendix}

\end{document}